\documentclass[fleqn,10pt]{wlscirep}
\usepackage[utf8]{inputenc}
\usepackage[T1]{fontenc}

\usepackage{lineno}
\usepackage{booktabs}
\usepackage{multirow}
\usepackage{longtable}
\usepackage[charter,cal]{mathdesign}

\title{The scales of urban mobility shape socioeconomic mixing}

\author[a,b,*]{Andrew Renninger}
\author[c,d,e]{Egor Kotov} 
\author[f]{Carmen Cabrera}

\affil[a]{Department of Network and Data Science, Central European University}
\affil[b]{School of Geographical \& Earth Sciences, University of Glasgow}
\affil[c]{Department of Spatial Planning, TU Dortmund University}
\affil[d]{Political and Social Sciences Department, Universitat Pompeu Fabra}
\affil[e]{Digital and Computational Demography Department, Max Planck Institute for Demographic Research}
\affil[f]{Geographic Data Science Lab, Liverpool University}

\affil[*]{Corresponding author: renningera@ceu.edu}

\begin{abstract}
Urban daily mobility determines how far we travel and whom we encounter through the city. As local-living and polycentric planning gain popularity, a key question is whether more localised mobility can preserve the socioeconomic mixing that cities enable. Yet systematic evidence on how different scales of movement contribute to experienced social exposure remains limited. Here, we use anonymised GPS traces from $\sim20$ million devices across 226 functional urban areas in France, Germany and the UK to decompose everyday mobility into three scales---proximal, medial and distal---and link these to the socioeconomic composition of destinations. Proximal trips remain similar in length across neighbourhoods, whereas medial and distal trips lengthen toward the periphery. Distal trips account for 33\% of travel but facilitate 42\% of inter-group encounters; reallocating them to local movement reduces experienced mixing by up to one third. These findings reveal a trade-off between shorter travel and integrated urban living.
\end{abstract}
\begin{document}

\flushbottom
\maketitle

\section*{Introduction}

With most of the world's socioeconomic production and innovation concentrated in urban areas \cite{glaeser2011triumph}, understanding and addressing the factors that contribute to the relative function or dysfunction of cities is of vital importance to continued urban sustainability. Among those factors, urban structure, and the way it shapes how people move through the city, has become a major concern in both research and policy. A core question is whether cities function better as monocentric systems organised around a dominant centre or as polycentric systems structured around multiple centres \cite{prud1999size, bertaud2021order}. Cities with a single centre can concentrate jobs and amenities, thicken labour markets, improve matching between workers and firms and reduce the infrastructure needed to connect people and activity \cite{andersson2007cities, combes2012productivity}. But a crowded centre can also result in increased congestion, slower journeys and higher housing costs near the centre \cite{ahlfeldt2019economic, couture2018speed, combes2019costs}. A polycentric structure can ease those pressures by distributing activity across secondary centres closer to where people live \cite{baum2007did, saiz2010geographic, duranton2004micro}, potentially improving accessibility \cite{ananat2011wrong, harari2020cities}, shortening travel times and reducing emissions \cite{mcdonald1987identification, giuliano1991subcenters, kloosterman2001polycentric}. This more decentralised arrangement underpins the ``15-minute city'' model, in which most daily needs can be met within a short walk, cycle or transit ride of home and which now features prominently in urban sustainability agendas \cite{bruno2024universal, calafiore2022twenty, zhang2025xminute, mouratidis2024challenge}.

Yet bringing daily life closer to home may carry a social cost. When residents meet most needs locally, they travel less across the wider city, and the encounters between people from different neighbourhoods and socioeconomic backgrounds that shared centres facilitate become less frequent \cite{abbiasov2024}. Localised living could narrow the socioeconomic range each resident is exposed to, even though such exposure helps cities widen opportunity, circulate information and maintain cohesion across social groups \cite{schlueter2010}. Spatial proximity enables social connections \cite{bailey2018social}, and the connections people form shape their economic prospects, by expanding access to knowledge and opportunities that are unevenly distributed across urban space \cite{granovetter1973strength, bailey2020social, chetty2022social}. Cities therefore face a fundamental tension between reducing travel by localising daily life and preserving the cross-neighbourhood interactions that integrate urban society. That tension is becoming more urgent as digital technologies reduce the need for physical co-presence, enabling remote work, online consumption and other substitutes for in-person interaction that once drew different groups into shared urban space \cite{dingel2020jobs, cabrera26latin}.

Despite the prominence of debates on how urban structure shapes socioeconomic mixing \cite{meijers2008measuring}, we still lack systematic evidence on how the spatial organisation of cities translates into the patterns of exposure residents experience in daily life. Traditionally, both urban structure and exposure to socioeconomic diversity have been studied from a static perspective, using measures of land use, employment density, remote sensing \cite{chen2017new, mcmillen2003number} and residential segregation \cite{cagney_urban_2020, muurisepp_activity_2022, liao_socio-spatial_2025, li_new-paradigm_2022}. Such measures show where activities and groups are located \cite{burger2012form}, but not which groups generate urban flows or how they encounter one another through daily movement. Approaches based on activity spaces show instead that socioeconomic mixing is dynamic and varies across times and urban locations, because daily mobility reshapes potential exposure to other groups \cite{xian_beyond-home_2022, zhang_mobility-based_2025, sun_activity-types_2024, jarv_ethnic_2015}. A significant body of work based on digital traces now demonstrates the potential of location data to capture these activity spaces \cite{xu2025experienced}, yet the evidence remains narrow, relying either on mobility data from a small number of cities \cite{roth2011structure, cabrera2023inferring} or on studies confined to a single country \cite{louail2014mobile, louail2015uncovering, xu2023urban, graells2021city}, and only rarely linking spatiotemporal patterns of movement to social outcomes. Research on experienced socioeconomic mixing has likewise focused mostly on selected American cities \cite{moro2021mobility, athey21, nilforoshan2023, renninger2025, wang2018urban, candipan2021residence}, and to a lesser extent, on European cities \cite{leroux2017clock, rossi_mori_time-space_2025, muurisepp_spatial-integration_2025}. Yet, we are still missing a harmonised comparison across a large and diverse urban sample that links explicit functional urban structure to experienced socioeconomic mixing.

Here we use anonymised GPS traces from $\sim20$ million devices across all 226 functional urban areas of France, Germany and the United Kingdom, to analyse how the everyday urban movement relates to socioeconomic mixing. We conceptualise urban structure as the composite outcome of multiple mobility behaviours associated with different urban functions, in line with the ideas of central place theory \cite{christaller1966central}, which posits a nested hierarchy of urban centres, each serving different mixes of activities, where some goods and services must remain close to home because they are needed frequently, whereas others can lie farther away because they are required only occasionally. Spatial \cite{zahavi1980regularities, marchetti1994anthropological, schafer1998global} and social constraints \cite{alessandretti2018evidence} place limits on how many places people can visit in a day, so urban structure emerges as businesses, populations and infrastructure adjust to these patterned demands. Recent empirical work shows that this logic operates both within cities and across systems of cities, with mobility organised across characteristic spatial scales: close and frequent, far and infrequent \cite{alessandretti2020scales, schlapfer2021universal}. Just as restaurants cluster into dining districts \cite{leonardi2023agglomeration}, people sort both between and within cities according to economic opportunity \cite{couture2020urban} and to the spatial distribution of amenities \cite{gaigne2022lives}.

Therefore, we conceptualise mobility as a multi-scale phenomenon and ask three questions: 
$\bullet$ Can everyday urban mobility be decomposed into interpretable spatial scales?\\
$\bullet$ How do these mobility scales vary across neighbourhoods and cities, and what do they reveal about monocentric versus polycentric urban structure?\\
$\bullet$ How do different mobility scales contribute to socioeconomic mixing?

In addressing these questions, we innovate in three ways. Methodologically, we introduce an unsupervised learning method for recovering the full range of spatial scales at which daily life is organised rather than a single characteristic distance based on previous work from Cabrera \emph{et al.} \cite{cabrera2023inferring}. This approach learns the structure from the data rather than imposing it \emph{a priori} \cite{batty2013new}. Empirically, we examine systems of cities orders of magnitude larger than prior studies, spanning three countries that differ in transport, urban form and welfare regime. Substantively, we trace those scales through to the socioeconomic mixing they produce, linking the structure of movement to a social consequence that bears directly on how sustainable and integrated a city can be.

\section*{Data and methods}
\subsection*{Mobility data}

We use anonymised GPS mobility traces from $\sim20$ million mobile devices across all 226 functional urban areas in France, Germany and the United Kingdom, observed over May and June 2024. Unlike transit smart-card records that capture only public transport \cite{cabrera2023inferring}, smart-phone traces register a rich variety of trips---errands near home that need no transit, and longer journeys that need a car---revealing the full spectrum of urban movement. For each device, we take the home location to be the modal \texttt{H3} level 10 cell occupied between midnight and 6:00 AM \cite{h3geo}. We apply the infostop algorithm \cite{aslak2020infostop} to detect stationary periods and extract discrete visits from continuous trajectories, where a visit is a spatially clustered set of GPS points persisting for at least 5 minutes within a level 10 cell. Each visit generates a trip of length $d_{ij}$, the haversine distance between the home origin $i$ and the destination $j$. Each trip is therefore anchored to home, because our aim is to characterise how residents of different neighbourhoods reach the wider city from where they live; we do not chain trips together.

This gives us $\sim200$ million trips across all cities. London has the most, at $\sim50$ million, and is the largest city, with a population near $\sim12$ million across its functional urban area. The trips form a spatial interaction network, which we illustrate for London, Paris and Berlin in Fig.~\ref{fig1}. As shown in Supplementary Fig.~\ref{si_validation_pop}, we find no detectable bias in the sample, based on the fact that the number of devices in a neighbourhood follows its population and, crucially, deviations do not vary with income. Coverage is similarly flat against the built environment, which Supplementary Fig.~\ref{si_validation_env} bears out.

\begin{figure*}[!ht]
\centering
\includegraphics[width=1\textwidth]{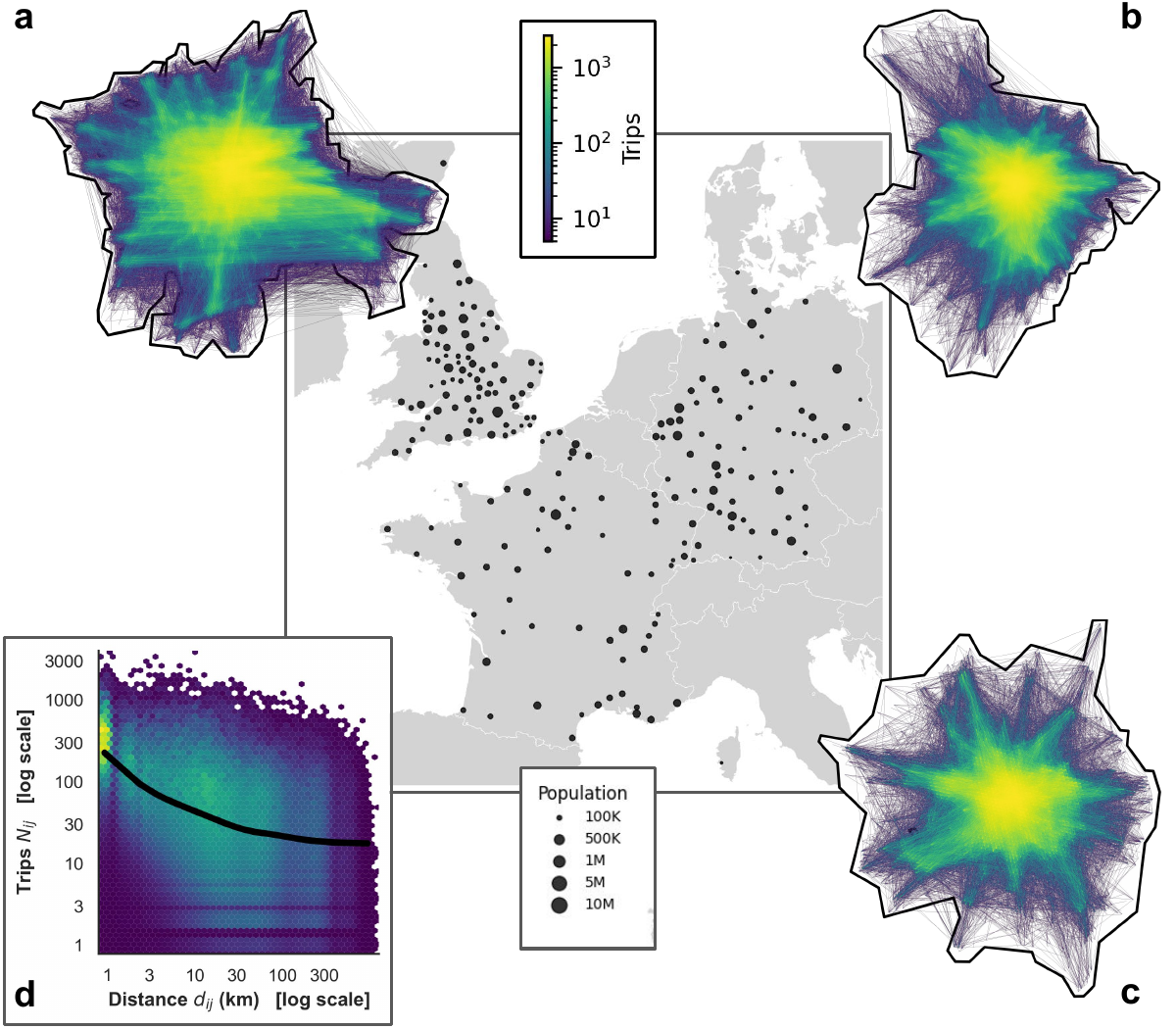}
\caption{\textbf{Everyday movement across 226 European cities.} The central map marks all 226 functional urban areas in the study, from towns of $\sim10^5$ residents to cities above $\sim10^7$. \textbf{a--c} Spatial interaction networks for London, Paris and Berlin; all three show strong core--periphery structure, and London in particular reveals smaller secondary hubs. \textbf{d} Aggregate distance--decay for London, relating trip length $d_{ij}$ to the number of journeys $N_{ij}$; the curve is a smoothed average of the binned data, and trip intensity falls gradually as distance grows.}
\label{fig1}
\end{figure*}

\subsection*{Socioeconomic data}
To see whom residents encounter, we attach a measure of socioeconomic standing to every neighbourhood and every destination. In England, we use area deprivation from the 2019 Index of Multiple Deprivation \cite{imd2019}; in France, median income per capita from INSEE's Filosofi \cite{filosofi2019}; and in Germany, average net rent per square metre from the 2022 Zensus \cite{zensus2022}. We harmonise these measures onto \texttt{H3} level 7 neighbourhoods and, because the three measures are not directly comparable, we sort neighbourhoods into five quintiles within each city, ordered so that the first quintile is the least advantaged and the fifth the most. We map these quintiles for nine case cities in Supplementary Fig.~\ref{si_income}. We run the mixing analysis on nine cities chosen to vary in size within each country---London, Birmingham and Manchester; Paris, Lyon and Marseille; Berlin, Hamburg and Munich. We read a quintile as relative standing within a city rather than income in levels, and we treat a visit to a higher-quintile destination as exposure to a different social setting, not as contact measured directly. Full sources and the harmonisation are given in Supplementary Section~\ref{si_data}.

\subsection*{Modelling the scales of mobility}
We infer the scales of urban mobility from the distribution of trip lengths in each neighbourhood. We group devices by home location, using \texttt{H3} level 7 cells---roughly five kilometres across---so that each model has enough trips, and we call these cells neighbourhoods. For each neighbourhood we collect the lengths of its trips, $\mathbf{l}_i=\{l_{i1},l_{i2},\dots,l_{iN_i}\}$, discarding all other information, and study their empirical distribution.

That distribution mixes errands near home with longer commutes and, as the example in Supplementary Fig.~\ref{si_multimodal} shows, it is often multimodal, suggesting that movement is organised at more than one scale. If a city were truly monocentric, with all goods and services gathered in a city centre, which we call the nucleus, nearly every trip would approximate the distance between the neighbourhood and that centre, so that $\mathbb{E}[L_i]\approx r_i$, the radial distance to the centre. A more polycentric or dispersed city spreads trips across several characteristic distances instead. To recover those distances without imposing them, and following \cite{cabrera2023inferring}, we model each neighbourhood's trip-length distribution as a finite mixture of Gamma distributions,

\begin{equation}\label{mixture}
f_i(l)=\sum_{c\in\{p,m,d\}}\pi_{ic}\,
\Gamma\!\bigl(l\,\bigl|\,\alpha_{ic},\beta_{ic}\bigr),
\tag{1}
\end{equation}

\noindent
where the weights satisfy $\sum_c\pi_{ic}=1$ and $\Gamma(l\mid\alpha,\beta)$ is the Gamma density with shape $\alpha$ and rate $\beta$. Trip lengths are positive and right-skewed, so we need a kernel whose support is $[0,\infty)$ and whose shape can flex from a bell to an exponential; the Gamma kernel nests the exponential ($\alpha=1$) and approaches the log-normal for larger $\alpha$, making it flexible yet tractable.

The three components $c\in\{p,m,d\}$ stand for proximal, medial and distal scales, which correspond, respectively to short trips within the surroundings of home, intermediate trips to nearby amenities such as shopping districts and longer trips to more distant destinations such as the workplace. The fitted model captures neighbourhood's mobility as six numbers, a mean and a weight for each of the three scales, $(\mu_{ic},\pi_{ic})$. We choose three components because they balance fit against interpretation. Across alternatives compared by the Bayesian Information Criterion, three components minimise the criterion in 90\% of neighbourhoods. We provide this comparison in Supplementary Fig.~\ref{si_model_selection} and Section~\ref{si_bic}.

We estimate the mixture for each neighbourhood with the Expectation--Maximisation algorithm, initialised by $k$-means clustering on log trip lengths, which stabilises the wide span of distances and gives plausible starting groups for near, middle and far trips. The algorithm alternates between assigning each trip a probability of belonging to each scale and updating each scale's parameters to match, until the log-likelihood settles; we give the full procedure in Supplementary Section~\ref{si_em}. After convergence we order the components by increasing mean, $\hat{\mu}_{ic}=\alpha_{ic}/\beta_{ic}$, and label them proximal, medial and distal, so that the components are read the same way in every neighbourhood. Because the model always fits three components, even a neighbourhood near the centre is assigned a distal component; there it captures trips \emph{away} from the nucleus rather than toward it. We account for this in the analysis by reading the weights---occasional leisure trips outward carry low weight---alongside the kinds of places the component's trips reach.

\subsection*{Relating the scales of mobility to urban structure}
The scales connect naturally to the shape of a city. In a monocentric city, trips from a neighbourhood point mainly at the single nucleus, so their lengths cluster around the radial distance $r_i$ to it; in a polycentric or dispersed city, trips spread across several centres and several characteristic distances. Mapped against distance from the centre, trip lengths fall well short of this monocentric benchmark in every city, as Supplementary Fig.~\ref{si_centrality_fig} shows. To make this concrete, we relate each component mean to a neighbourhood's position in the city. Following \cite{cabrera2023inferring}, we define a city's effective centre, or nucleus, as the centroid of the hexagon with most visits during our study period, let $r_i$ be the distance from that centre to neighbourhood $i$, and estimate

\begin{equation}\label{structure}
\hat{\mu}_{ic}=a_c + b_c\,r_i + \varepsilon_{ic},
\tag{2}
\end{equation}

\noindent
by weighted least squares, with the trip count $N_i$ as weights. The slope $b_c$ measures how sharply each scale responds to the centre--periphery gradient, and the intercept $a_c$ measures its reach for a resident at the centre itself. In a polycentric city we expect $b_p\!\approx\!0$, as local activity stays local everywhere, with $b_m\!>\!0$ and $b_d\!>\!b_m$ as residents reach toward central places for amenities and work.

\subsection*{Quantifying socioeconomic mixing across scales}
We then ask how each scale exposes residents to other socioeconomic groups. Using the fitted model, we assign every observed trip to its most likely scale and we record the socioeconomic quintile of each destination it reaches. For a neighbourhood $i$ and scale $c$, the destinations reached describe a distribution over the five quintiles; we summarise the breadth of that exposure with the Hill number of order 1, the exponential of its Shannon entropy, which we read as the effective number of socioeconomic groups a resident encounters through scale $c$. Higher values mean broader and more even exposure. We report Hill numbers of orders 0 and 2 and an evenness ratio for robustness, and we summarise city-wide patterns with origin--destination interaction matrices by quintile; all are defined in Supplementary Section~\ref{si_mixing}.

Finally, we ask whether long trips are necessary for mixing, using counterfactual mobility regimes that hold each neighbourhood's trip count and fitted distances fixed but change where trips go, which we detail in Supplementary Section~\ref{si_counterfactual}. The first is a null that sends trips to random reachable destinations at the same distances, isolating the role of destination \emph{choice} from that of distance. The other two remove the distal component and reassign its trips to shorter scales---in one case to the medial scale, in the other to the proximal---to simulate a city that has shed its longest journeys, as a broad move to remote work or a thoroughgoing 15-minute city might. For each regime we recompute the interaction matrices and Hill numbers and report the change in city-level mixing relative to what we observe.

\section*{Results}
\subsection*{Everyday mobility resolves into three scales across European cities}
\begin{figure*}[!ht]
\centering
\includegraphics[width=0.9\textwidth]{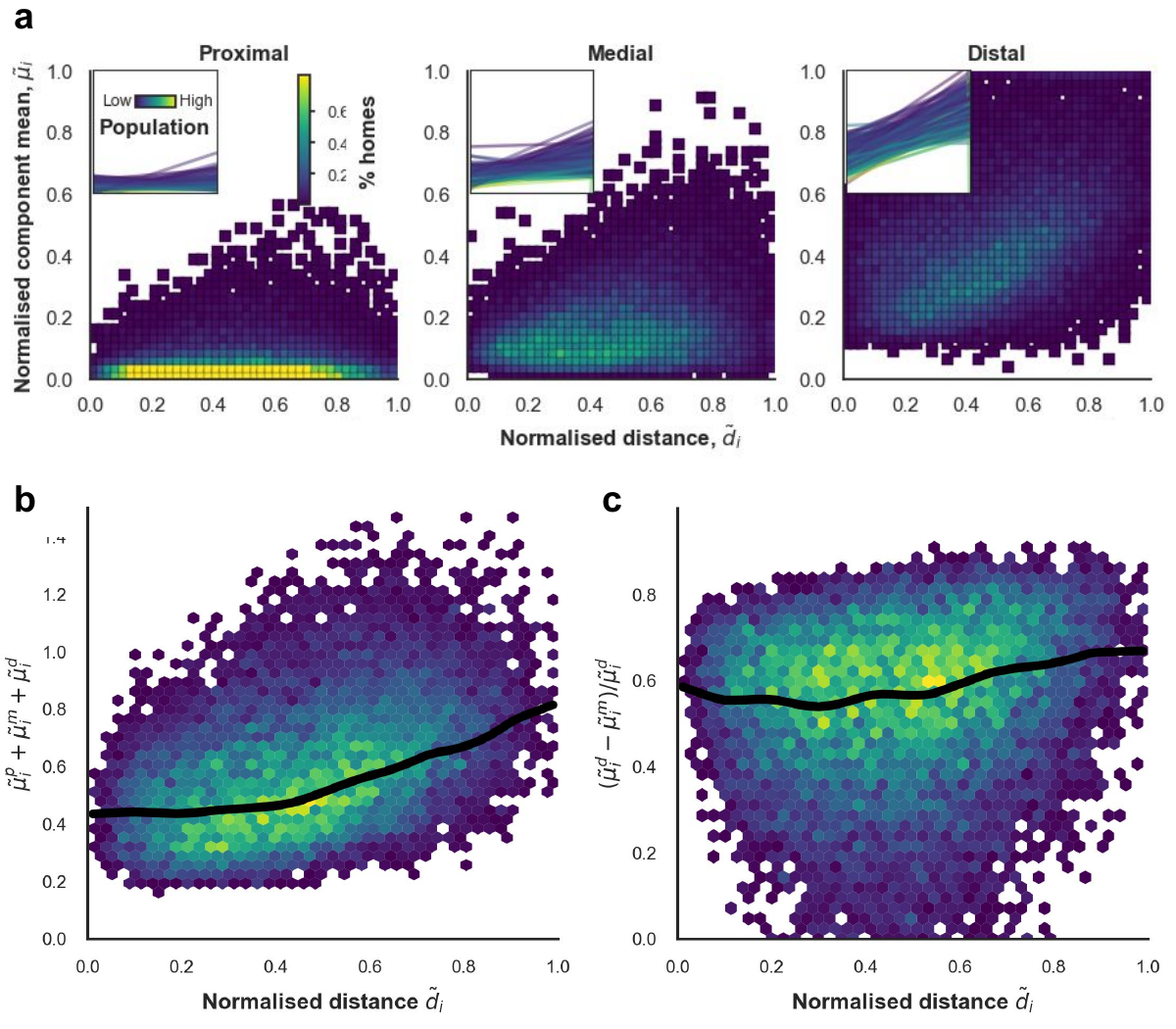}
\caption{\textbf{Mobility resolves into three scales across European cities.} \textbf{a} Each component mean against normalised distance from the city centre, pooling 226 cities. The proximal scale is flat, while the medial and distal scales rise toward the periphery and the distal rises most steeply. Insets show the fitted line for each city, coloured by population rank: the gradient is steeper in smaller cities, which lean harder on their core. \textbf{b} The weighted sum of the three scales---the total distance a neighbourhood's residents must cover---rises with distance from the centre, so peripheral residents travel farther for the same day. \textbf{c} The gap between distal and medial trip lengths widens toward the edge, as the longest journeys grow more distinct from intermediate ones.}
\label{fig4}
\end{figure*}

We begin with the system as a whole, fitting the three-component mixture to roughly 20,000 neighbourhoods across roughly 200 cities. To compare cities of very different size, we normalise each so that the longest distance from its centre equals one. Fig.~\ref{fig4}\textbf{a} plots each component mean against this normalised distance, pooling neighbourhoods across all cities, with the insets showing the fitted line for each city coloured by population rank.

Three regularities stand out. First, he proximal mean is nearly flat, suggesting that local life operates at the same scale whether a neighbourhood sits at the centre or the edge, and whether the city is large or small. Second, the medial and distal means both rise toward the periphery, but the distal rises far more steeply, suggesting that longer journeys are tied to the centre--periphery gradient in a way that local movement is not. Third, as a result of these patterns, the total distance a neighbourhood's residents must cover, the weighted sum of the three means $D_i=\hat{\pi}_i^p\tilde{\mu}_i^p+\hat{\pi}_i^m\tilde{\mu}_i^m+\hat{\pi}_i^d\tilde{\mu}_i^d$, climbs with distance from the centre, traced by the curve in Fig.~\ref{fig4}\textbf{b}: peripheral residents pay a distance premium, travelling farther for the same day. Fig.~\ref{fig4}\textbf{c} adds that the gap between distal and medial trips widens toward the edge, so the longest journeys become increasingly distinct from intermediate ones the farther out a neighbourhood lies.

The composition of trips across the three scales is remarkably stable from city to city and country to country, a regularity Supplementary Fig.~\ref{si_variation} lays out in full. What does vary, modestly, is the pull of the centre. In larger cities the distal mean rises less steeply with distance and from a lower base, a sign that growth loosens a city's reliance on a single nucleus and spreads activity across secondary centres. The relationship is noisy and we return to it with the case cities below.

\subsection*{Mobility scales reveal variations in urban structure across cities}

We further examine neighbourhood-level variation in mobility scales and what this variation reveals about urban structure. To aid interpretation of the Europe-wide regularities identified in Fig. \ref{fig4}, we focus on three case studies corresponding to London, Paris and Berlin. For each city, we relate the proximal, medial and distal components of each neighbourhood to the neighbourhood's relative location within the urban system. This allows us to assess whether the observed mobility scales are more consistent with a monocentric organisation around a dominant nucleus or with a more diffuse, polycentric urban form.

\begin{figure*}[!ht]
\centering
\includegraphics[width=0.9\textwidth]{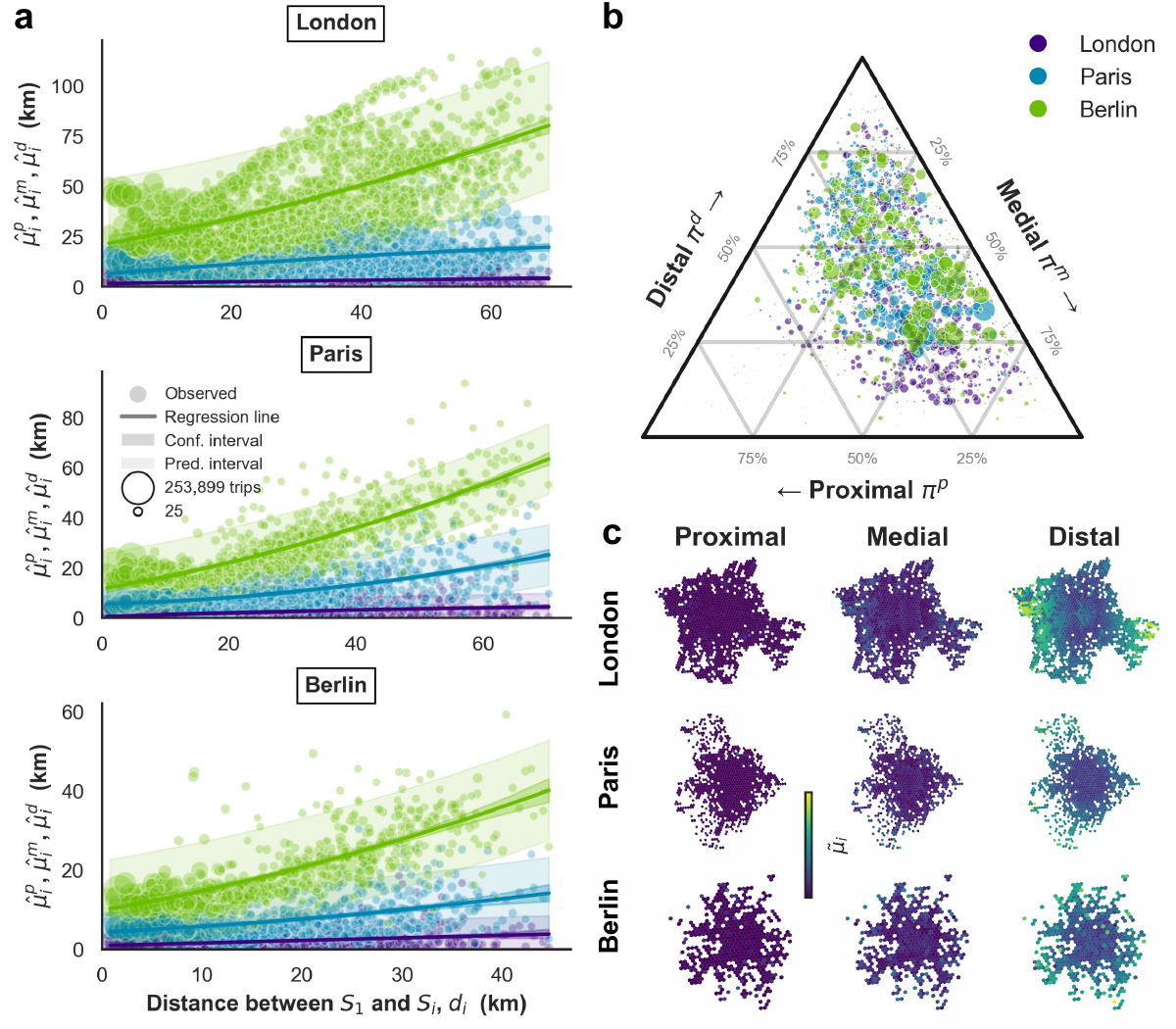}
\caption{\textbf{The scales trace each city's structure.} \textbf{a} Proximal, medial, and distal component means against distance from the centre for London, Paris, and Berlin; the proximal scale is flat in all three, while the distal scale rises toward the periphery. \textbf{b} The component weights---the share of trips at each scale---for each neighbourhood; London carries higher proximal and medial weights, the mark of a more polycentric city. \textbf{c} Maps of the three component means, showing low means across scales at the centre and high distal means at the periphery.}
\label{fig3}
\end{figure*}

The three cities share the same three mobility scales, but differ in how those scales are organised spatially. Fig.~\ref{fig3}\textbf{a} shows proximal trips varying little with distance in all three. Local life runs at a common scale regardless of where, or which city, a neighbourhood sits in. Distal trips, by contrast, lengthen sharply toward the periphery, the signature of a core--periphery structure in which city-wide journeys matter more the farther out one lives---and the maps in Fig.~\ref{fig3}\textbf{c} carry the same gradient, with distal means highest at the edge. The largest cities are not special in this, as we find similar, and sometimes starker, gradients in other smaller European cities, which Supplementary Fig.~\ref{si_component_scatter_cities} sets out for the second- and third-largest in each country. The clearest difference between the cities is in the weights and Fig.~\ref{fig3}\textbf{b} sets London apart. London carries higher proximal and medial weights and a more dispersed mix of scales than Paris or Berlin, meaning more of daily life there is met locally or at intermediate distance rather than through long trips to the core. London, often called a collection of villages, looks more polycentric than Paris and Berlin, which lean more on a dominant centre. The same model assigns central neighbourhoods a distal component that points outward rather than in; in those neighbourhoods distal trips are long but carry little weight and they sit alongside poorer model fit at the city's edge. We map these patterns in Supplementary Figs.~\ref{si_weights} and~\ref{si_residuals}.

\subsection*{Long journeys drive social mixing}

We next examine the extent to which the different mobility scales identified above contribute to social mixing. To do so, we assign both origins and destinations to one of five socioeconomic quintiles using country-specific small-area indicators of relative socioeconomic advantage and disadvantage. These are area-level deprivation in England, income in France and rent in Germany. We then assess how proximal, medial and distal trips expose residents to different degrees of socioeconomic diversity. 

\begin{figure*}[!ht]
\centering
\includegraphics[width=0.9\textwidth]{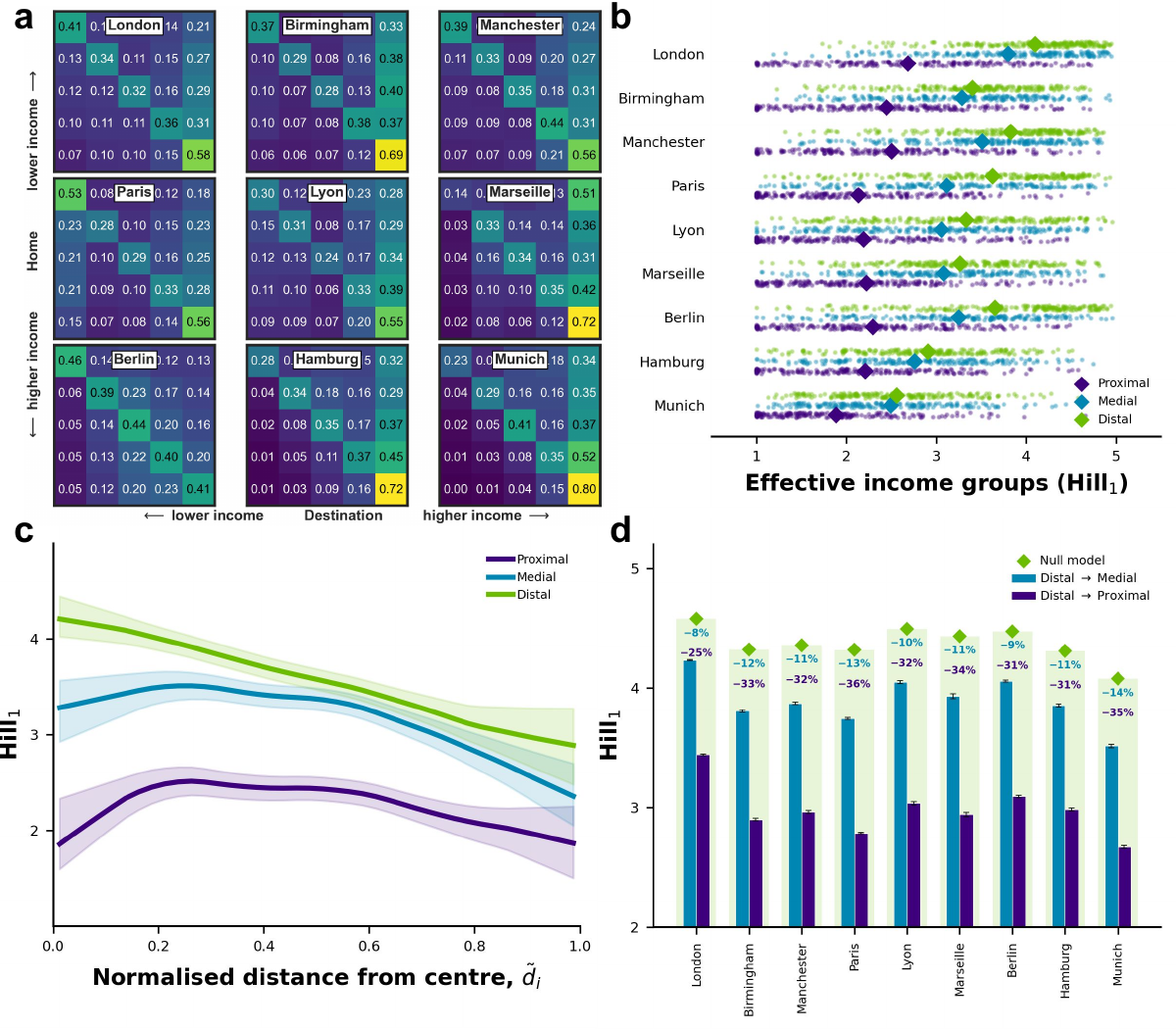}
\caption{\textbf{Long journeys carry social mixing.} \textbf{a} Interaction matrices giving the probability $P(j \mid i)$ that a resident of socioeconomic quintile $i$ visits a destination in quintile $j$, for nine cities across England, France and Germany; the diagonal captures homophily, and the upper-right mass a tendency for residents of lower-status areas to visit higher-status destinations. \textbf{b} The effective number of socioeconomic groups encountered (Hill$_1$) across neighbourhoods, decomposed by scale; distal trips are the most diverse, proximal the least, with the median marked. \textbf{c} Mixing against normalised distance from the centre, pooled across the nine cities, by scale. \textbf{d} Change in mixing under two regimes that remove the distal component: reassigning distal trips to the medial scale lowers Hill$_1$ by 8--14\%, while reassigning them to the proximal scale alone lowers it by 25--36\%.}
\label{fig6}
\end{figure*}

The interaction matrices in Fig.~\ref{fig6}\textbf{a} give the probability that a resident of one socioeconomic quintile visits a destination in another. Two patterns recur in every city. Movement is homophilic, in the sense that most interactions sit on the diagonal, suggesting that like visiting like. The off-diagonal mass is asymmetric, with residents of lower-status neighbourhoods reaching higher-status destinations more often than the reverse. This tendency holds across the full set of matrices in Supplementary Figs.~\ref{si_matrices_eng}--\ref{si_matrices_ger}. This poorer-to-richer pull is consistent with the gradient of land rent that places wealthier residents and the amenities they support near the core, and it echoes a poorer-to-richer asymmetry already seen in urban consumption and mobility \cite{dong2020segregated, hilman2022socioeconomic, bokanyi2021universal, renninger2025}. To our knowledge, though, it has not before been documented across three European countries with such different urban and social structures.

Decomposing this exposure by scale shows where the mixing comes from. The distribution in Fig.~\ref{fig6}\textbf{b} ranks the scales. Distal trips produce the highest socioeconomic diversity, proximal the lowest, and medial sits in between. Reading mixing against distance, Fig.~\ref{fig6}\textbf{c} finds it higher for neighbourhoods nearer the centre at every scale. The distal scale stays more diverse than the proximal across the whole gradient. This pattern repeats across cities in Supplementary Fig.~\ref{si_distance_decay_fig}. The imbalance is striking: distal trips are about a third of all travel, yet they carry 42\% of the encounters between different socioeconomic groups. Long journeys are critical the mixing residents experience.

A counterfactual analysis in Fig.~\ref{fig6}\textbf{d} helps us understand this dependence on long journeys precise. We first separate the contributions of distance choice by holding each neighbourhood's trips and their lengths fixed but sending them to random reachable destinations; this null model mixes residents \emph{more} than they actually mix---in every city. Across the nine cities, the null produces between 19\% and 47\% more socioeconomic mixing than we observe, reaching 19\% in London, 24\% in Paris and 29\% in Berlin. This suggest that residents travel with a bias: if they scattered across the city at random---which is an unrealistic but informative standard---they would mix far more. We then remove the distal component and reassign its trips to shorter scales. Sending distal trips to the medial scale lowers mixing by 8--14\%, around 11\% on average and most in Paris and Munich; sending them to the proximal scale only lowers it by 25--36\%, reaching 25\% in London, 36\% in Paris and 31\% in Berlin. Supplementary Fig.~\ref{si_counterfactual_fig} reports every city. Roughly a third of the socioeconomic variety residents encounter rides on their longest journeys, and the medial scale recovers most---though not all---of what the distal scale provides. Supplementary Fig.~\ref{si_mixing_2x2} brings these strands together for all nine cities. Distal trips carry a disproportionate share of contact across wealth brackets, even as residents mix below what their travelled distances would allow. Distal mobility widens the range of groups encountered without making exposure across them appreciably more even, so its advantage is one of scope rather than balance, as the diversity and evenness measures in Supplementary Fig.~\ref{si_diversity} confirm.

If mixing can be had efficiently at the medial scale, a city might route its longest trips to nearby subcentres rather than across town. We test this lever with a radiation model \cite{simini2012universal, noulas2012tale}, which assigns trips according to both proximity and opportunity, or the number of amenities, on the premise that well-placed ``hubs'' can bridge groups that would otherwise stay apart \cite{nilforoshan2023}. Because hubs beyond the nucleus exist in the visitation data, we choose subcentres from them---the local maxima of the visit surface, kept at least five kilometres apart to ensure that they are discrete clusters. Assigning distal trips to these medial nodes shortens the average journey by about 45\%, but if a neighbourhood's residents visit the same subcentres consistently and repeatedly, according to their size and closeness, it still hardens segregation, as Supplementary Fig.~\ref{si_medialisation}\textbf{a} shows the mixing slipping below the diffuse medial trips they already make. Subcentres typically exist in the wealthier parts of the city, as Supplementary Fig.~\ref{si_medialisation}\textbf{b} shows, so visiting them raises how far up residents reach while narrowing the social range of the places they reach---more reach, less variety, in seven of the nine cities in Supplementary Fig.~\ref{si_medialisation}\textbf{c}. That is, if visitors show a strong tendency to return to a small set of subcentres at the medial length, experienced segregation remains high, but if they explore more it falls. Across the subcentre definitions in Supplementary Fig.~\ref{si_medialisation}\textbf{d}, the gap to the diffuse medial closes only as the set of hubs grows and the smallest sets fall below even the local floor.

\section*{Discussion}
Cities are often considered more sustainable when daily needs can be met closer to home, replacing long commutes with shorter local trips. Our results add a social entry to that account. The trips a compact, local city would shed are the same trips that carry its residents into contact with diverse people from different socioeconomic backgrounds. Across nine cities, the longest journeys constitute about a third of travel but produce close to half of all cross-group encounters, and a city that reassigned those journeys to local movement would lose between a tenth and a third of the socioeconomic mixing its residents now experience. Bringing life closer to home narrows the social range of the day, as mobility beyond local residential environments expands the range of groups reached \cite{moro2021mobility, chen_how-far_2025}. Longer journeys break the spatial bounds of residential sorting, widening the potential exposure to diverse social groups and allowing cross-neighbourhood encounters that local movement naturally constrains. 

Yet travel distance by itself does not guarantee social mixing. If daily urban travel replicates residential isolation across other activity locations, it feeds into the ``vicious circles of segregation'' \cite{tammaru_spatial_2021, van_ham_urban_2021}. Socioeconomic mixing thus depends on how far residents travel and on how they sort across destinations. Holding distance fixed and sending residents to random destinations produces more mixing than we observe, in every city, so the limit on encounter is not only how far people can travel but where they choose to go. Residents use their longest trips to reach the centre, where different groups converge, but they also sort: for example, a poorer resident is more likely to travel into a wealthier district than the reverse. Distance is necessary for mixing, then, but not sufficient. Because long trips follow the bid-rent gradient, wherein the centre of a city is wealthier than its periphery because property values are high there, a process by which less wealth residents move to the centre each day also puts the burden of mixing on them. Moreover, proximity to daily needs bring them within reach, but it does not ensure residents use them locally, let alone meet a varied set of others there \cite{maciejewska2025proximity}.

This means that there is no obvious solution for cities. Leaning on the centre produces mixing, but it lengthens journeys. Peripheral residents already pay a distance premium, travelling farther for the same day. Moving activity to local scales cuts that premium but, as the counterfactuals show, hits mixing hardest. Between the two lies the medial scale---the intermediate trips to secondary hubs and shopping districts that recover most of the mixing without the longest journeys, the same secondary centres that already shorten the commute for those who live near them \cite{cervero1997polycentrism}. When we move distal trips onto the medial scale rather than the proximal one, the loss of mixing shrinks from about a third to roughly a tenth. The key question, then, is not whether a city is monocentric or polycentric in the abstract, but how much of daily life can be shifted onto this intermediate scale through planning that strengthens the secondary centres supporting medial travel.

But which secondary centres a city leans on is not incidental. Route long trips to a city's actual subcentres rather than spread them across the medial scale, and mixing slips rather than holds: subcentres sit in the wealthier parts of the city, so sending residents there raises the standing of the places they reach but narrows the social range of those places. The medial scale mixes through the variety of places it reaches, not by reaching higher---so the lever is not subcentres as such, but socioeconomically varied ones \cite{nilforoshan2023, juhasz2023amenity, fan2023diversity, kim_segregated-by-whom_2024}. Because advantaged groups can self-segregate by frequenting exclusive amenities or differentiated subcentres, functional polycentricity must be evaluated not merely by the physical layout of subcentres, but by whether the mobility flows bridge neighbourhoods or actively reproduce residential socioeconomic divisions.

Some of what we find is reassuringly stable. The proximal scale---the radius of genuinely local life---is nearly the same in every neighbourhood of every city, an empirical counterpart to the 15-minute neighbourhood that already exists, without policy \cite{ellder2024built}. What varies is everything above it, namely, how far residents must reach for the rest of the city and how much of that reach the centre commands. On this evidence the 15-minute city is less a structure to build than a share of activity to protect, while keeping the longer journeys that do the mixing.

Our evidence has limits. We observe movement and the places it reaches, not contact itself. Two residents in the same district on the same day may never meet, so our measures capture the opportunity for mixing rather than the mixing that occurs. We capture exposure---the chance of co-presence, not the ties it may seed or the welfare those ties carry \cite{chetty2022social}. Socioeconomic standing is relative within each city and read from area measures, not individuals. Our approach also aggregates journeys without distinguishing between work, routine errands and leisure, even though the context and timing of a trip shape the potential for mixing \cite{sun_activity-types_2024, muurisepp_spatial-integration_2025, muurisepp_activity_2022, liao_socio-spatial_2025}. Mixing relies heavily on discretionary and non-routine activities, whereas workplace and residential mobility may reflect different constraints and distinct degrees of socio-spatial isolation. And because advantage is spatially clustered, reaching a different group often requires distance, which could make the link between long trips and mixing partly mechanical; the null comparison guards against the simplest version of this concern, since observed mixing falls below a distance-matched benchmark, but a sharper test is whether the pattern weakens in cities where groups are less segregated to begin with---a check we leave for work that extends this one.

The appeal of the local city is real. Shorter trips, lighter demands on travel, time returned to residents. But a city is also a place where strangers of different means share a street, a train, a queue, and those encounters happen disproportionately on the journeys a local city is designed to remove. Sustainability framed only as distance may quietly trade away the mixing that makes a city more than the sum of its neighbourhoods. The harder task, and the one these measures are built to support, is to shorten the journey without shrinking the city.


\begingroup
\renewcommand{\addcontentsline}[3]{}
\bibliography{polycentricity}
\endgroup

\section*{Acknowledgements}
The authors thank \href{https://locomizer.com/}{Locomizer} for providing the mobility data used in this study. Access to the data was supported by a grant from the Centre for Digital Innovation at University College London.

\section*{Author contributions statement}
\textbf{A.R.} Conceptualisation, data access, methodology, analysis, writing, visualisation, project management. \textbf{E.K.} Conceptualisation, methodology, writing. \textbf{C.C.} Conceptualisation, methodology, writing.

\subsection*{Data and code availability}
Because of data protection rules, we are not able to share the mobility data used in this study. Aggregates that allow for reproduction of key analyses can be found at \href{https://github.com/asrenninger/polycentricity}{https://github.com/asrenninger/polycentricity}.

\section*{Competing interests}
The authors declare no conflict of interest.

\clearpage
\setcounter{section}{0}
\setcounter{figure}{0}
\setcounter{table}{0}
\setcounter{equation}{0}
\renewcommand{\thefigure}{S\arabic{figure}}
\renewcommand{\thetable}{S\arabic{table}}
\renewcommand{\theequation}{S\arabic{equation}}

\begin{center}
{\Large Supplementary Information for\\[6pt] \textbf{The scales of urban mobility shape socioeconomic mixing}}
\end{center}

\renewcommand{\contentsname}{Supplementary Notes}
\tableofcontents
\clearpage

\section{Coverage and representativeness of the mobility data}\label{si_validation}

\begin{figure*}[!ht]
\centering
\includegraphics[width=0.95\textwidth]{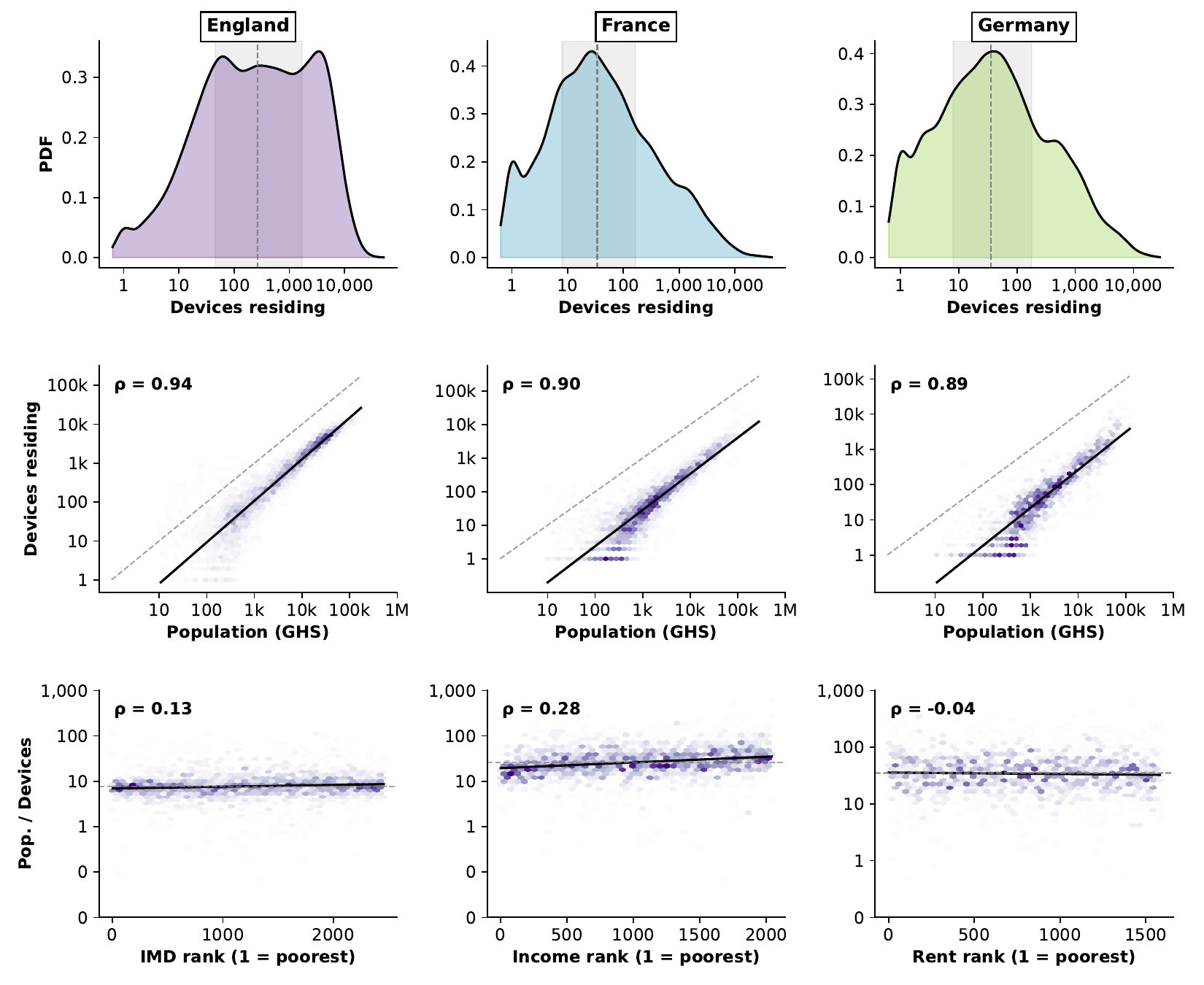}
\caption{\textbf{The device sample follows population, not income.} Columns are England, France, and Germany, each pooling its three case cities. \textbf{Top row}: the distribution of devices residing per neighbourhood, with the interquartile range shaded and the median marked. \textbf{Middle row}: devices against population (GHS), with a fitted line, the 1:1 line dashed, and Spearman's $\rho$; device counts track population. \textbf{Bottom row}: people per device against socioeconomic rank (1 = poorest), with Spearman's $\rho$; coverage shows little systematic gradient across the income distribution.}
\label{si_validation_pop}
\end{figure*}

Before we read structure into these traces, we check that the device sample reflects the cities it is meant to describe. Pooling the three case cities in each country, we ask whether the number of devices we observe in a neighbourhood tracks how many people live there, and whether coverage tilts toward richer or poorer areas. Supplementary Fig.~\ref{si_validation_pop} shows that it does the first and not the second: device counts rise closely with population, while the number of people per device stays roughly flat across the socioeconomic gradient, so the sample neither favours nor neglects the wealthy. We extend the same check to the built environment in Supplementary Fig.~\ref{si_validation_env}, relating coverage to building height, vegetation, and air quality; none of the three bends it appreciably. The traces are, in short, a broad and even sample of the population rather than a select slice of it.


\begin{figure*}[!ht]
\centering
\includegraphics[width=0.95\textwidth]{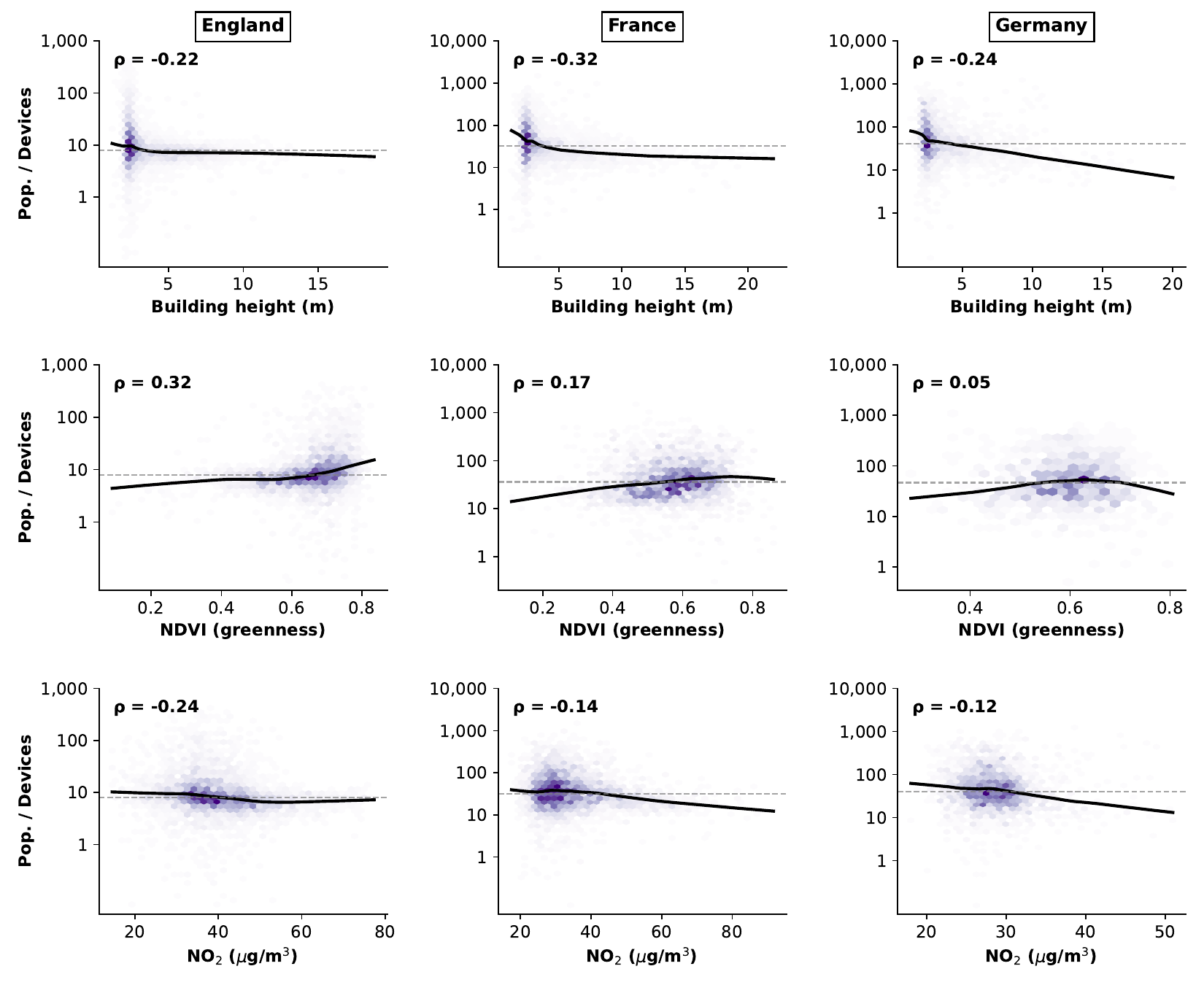}
\caption{\textbf{Coverage is flat across the built environment.} Columns are England, France, and Germany, each pooling its three case cities; rows relate people per device to building height, vegetation (NDVI), and nitrogen dioxide (NO$_2$). Coverage is broadly flat across all three, so the sample is not strongly tied to how built-up, green, or polluted a neighbourhood is.}
\label{si_validation_env}
\end{figure*}

\clearpage

\section{Socioeconomic data and harmonisation}\label{si_data}

We measure socioeconomic standing from the most authoritative national source in each country: area deprivation from the 2019 Index of Multiple Deprivation in England, median income per capita from INSEE's Filosofi in France, and average net rent per square metre from the 2022 Zensus in Germany. Because the three measures are not comparable in their native units, we sort neighbourhoods into five quintiles within each city, ordering them so that the first quintile is the least advantaged and the fifth the most. We attach a quintile to every home neighbourhood and to every destination, and we read it as relative standing within a city rather than as income in levels.

We map the resulting quintiles for the nine case cities in Supplementary Fig.~\ref{si_income}. In most cities the gradient runs from the periphery toward the core, so that reaching a different socioeconomic group is, in part, a matter of distance---a point we return to when we ask whether long trips drive mixing because of where people go or only because of how far they travel.


\begin{figure*}[!ht]
\centering
\includegraphics[width=0.85\textwidth]{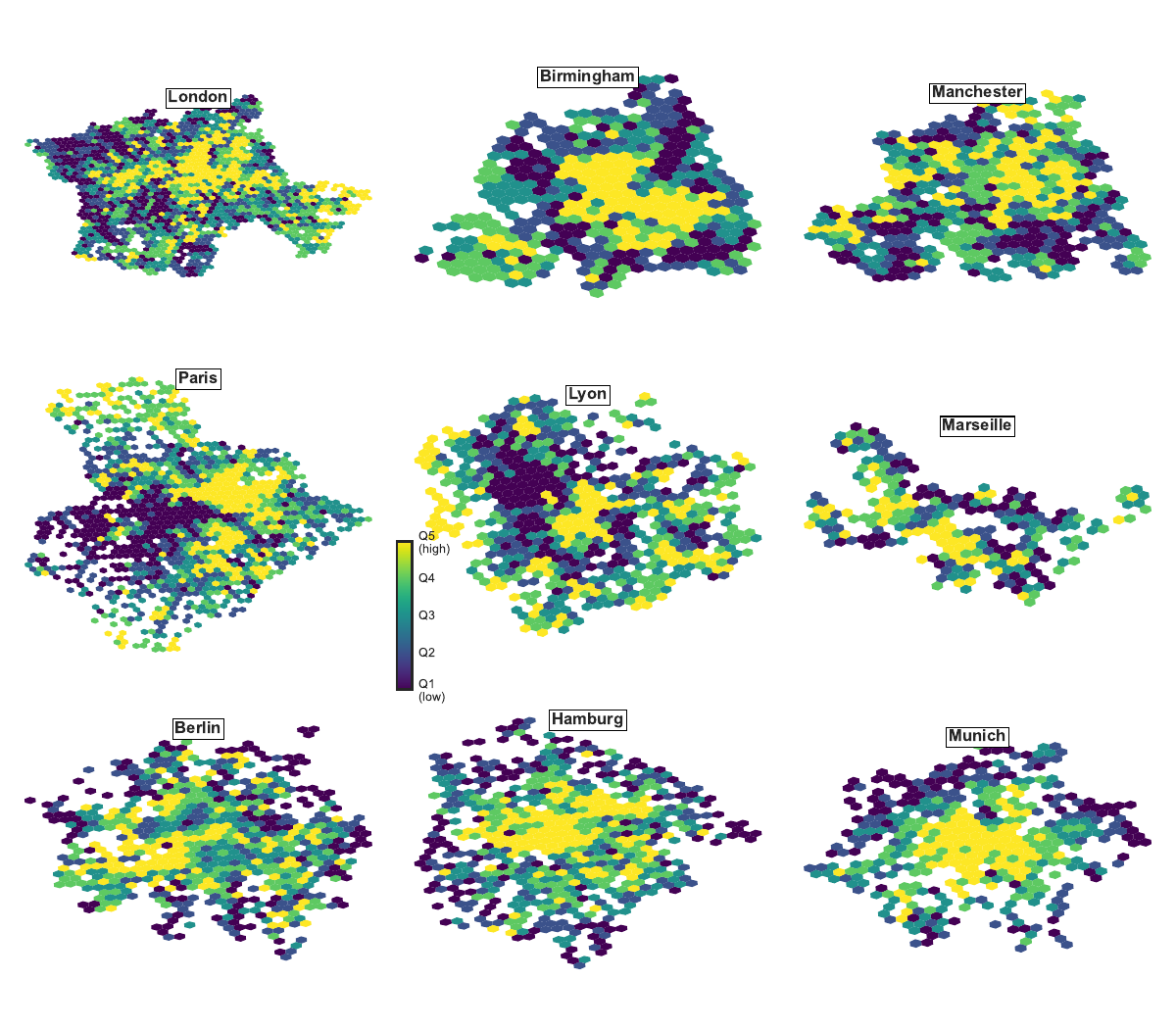}
\caption{\textbf{In most cities, standing rises from periphery to core.} Each map colours \texttt{H3} level 7 neighbourhoods by their within-city socioeconomic quintile, from the least advantaged (Q1) to the most (Q5). Rows group the cities by country---England, France, and Germany. Standing is measured from area deprivation in England, income in France, and rent in Germany, then sorted into quintiles separately within each city so that a quintile reads as relative position in that city.}
\label{si_income}
\end{figure*}

\clearpage

\section{Comparability of the socioeconomic indicators}\label{si_collinearity}

We rank neighbourhoods by deprivation in England, by income in France, and by rent in Germany, taking in each country the indicator that national statistics measure best and publish at the finest scale. A fair question follows: do these three order neighbourhoods the same way, or have we stitched together signals that pull apart? England lets us check directly, because there all three coexist at the neighbourhood scale. Alongside the deprivation score we already use, we draw two further small-area releases from the Office for National Statistics---net annual household income \cite{onsincome2023} and the median price paid for a home \cite{onshpssa2025}---and ask how closely they move together across the 6{,}677 middle-layer areas, some 7{,}800 residents each, where all three are published. Price stands in for rent here, since rents at so fine a scale are not openly published; the two move almost in step \cite{ahlfeldt2022micro}.

\begin{figure*}[!ht]
\centering
\includegraphics[width=0.95\textwidth]{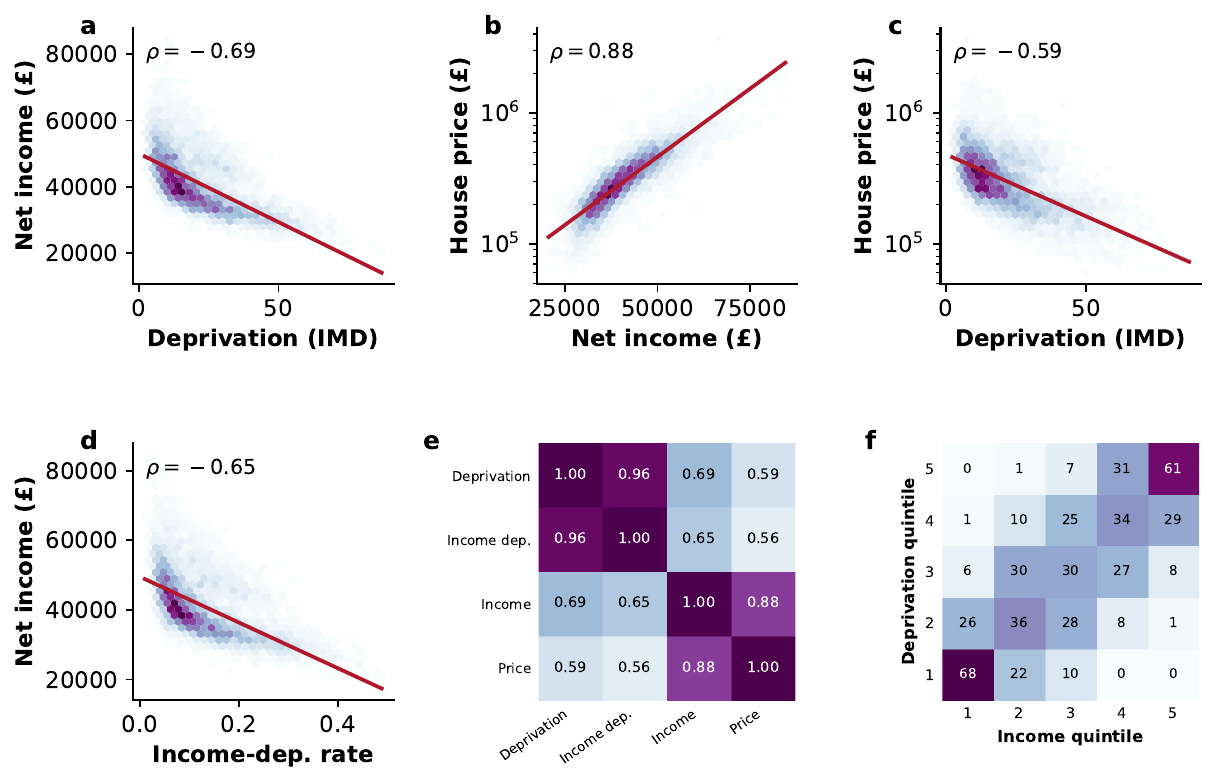}
\caption{\textbf{Deprivation, income, and housing cost track one another across English neighbourhoods.} Each point is a middle-layer super output area (MSOA, roughly 7{,}800 residents); we plot the 6{,}677 areas where all three signals are published. \textbf{a--c}, The three cross-country pairings---deprivation against income, income against housing cost, and deprivation against housing cost---each with a least-squares fit and the Spearman rank correlation $\rho$; housing cost is drawn on a logarithmic axis. \textbf{d}, Income measured two ways, the income-deprivation rate against survey net income. \textbf{e}, Spearman correlations among all four signals, each oriented so that higher means better-off. \textbf{f}, Within our study cities, the share of neighbourhoods in each income quintile (columns) by deprivation quintile (rows); the mass sits on the diagonal, and ninety per cent of neighbourhoods fall within one quintile whichever signal ranks them. Deprivation is the 2019 Index of Multiple Deprivation score and its income-domain rate (measured at lower-layer areas and population-weighted to MSOA); income is ONS net annual household income for the financial year ending 2023; housing cost is the ONS median price paid for the year ending September 2025.}
\label{si_collinearity_fig}
\end{figure*}

They move together closely. Supplementary Fig.~\ref{si_collinearity_fig} reads as one diagonal band repeated from panel to panel: a more deprived neighbourhood earns less and houses cost less, and the three orderings agree far more than they diverge. Income and housing cost track one another most tightly, at a Spearman correlation of $\rho = 0.88$; deprivation follows income at $\rho = 0.69$ and housing cost at $\rho = 0.59$; and the income a survey measures and the income deprivation an index counts agree at $\rho = 0.65$. Within a single city---the scale at which we cut our quintiles---the agreement only sharpens once national gradients are removed, with deprivation and income reaching $\rho = 0.79$ across London, Birmingham, and Manchester. The quintiles make the point plainest: sort a city's neighbourhoods into fifths by income or by deprivation, and the two rankings place ninety per cent of them within a single fifth of one another. Choosing income over deprivation, or rent over either, would reshuffle the edges of our quintiles and leave their substance intact.

That the three coincide is no accident of England. A neighbourhood's rent capitalises what its residents can pay to live there, so income and housing cost rise together wherever households bid for location, as Alonso set out and hedonic studies have measured since \cite{alonso1964location,rosen1974hedonic}. Britain's deprivation index builds income in by construction, as one of its domains \cite{noble2006measuring}, and the wider study of income segregation treats deprivation, earnings, and housing cost as facets of one gradient \cite{reardon2011income}. The same gradient organises the cities we study abroad: prices and rents move together across Germany down to the postcode \cite{ahlfeldt2022micro}, and a sharp social segregation runs through its cities \cite{helbig2018bruchig}; for France and its European neighbours, a long literature documents the same sorting of rich from poor across urban space \cite{tammaru2016socioeconomic,musterd2017socioeconomic}. We therefore read our three national indicators as three windows on one underlying position, and our within-city quintiles as comparable measures of relative standing.

\clearpage

\section{Multimodality in trip-length distributions}

\begin{figure*}[!ht]
\centering
\includegraphics[width=0.85\textwidth]{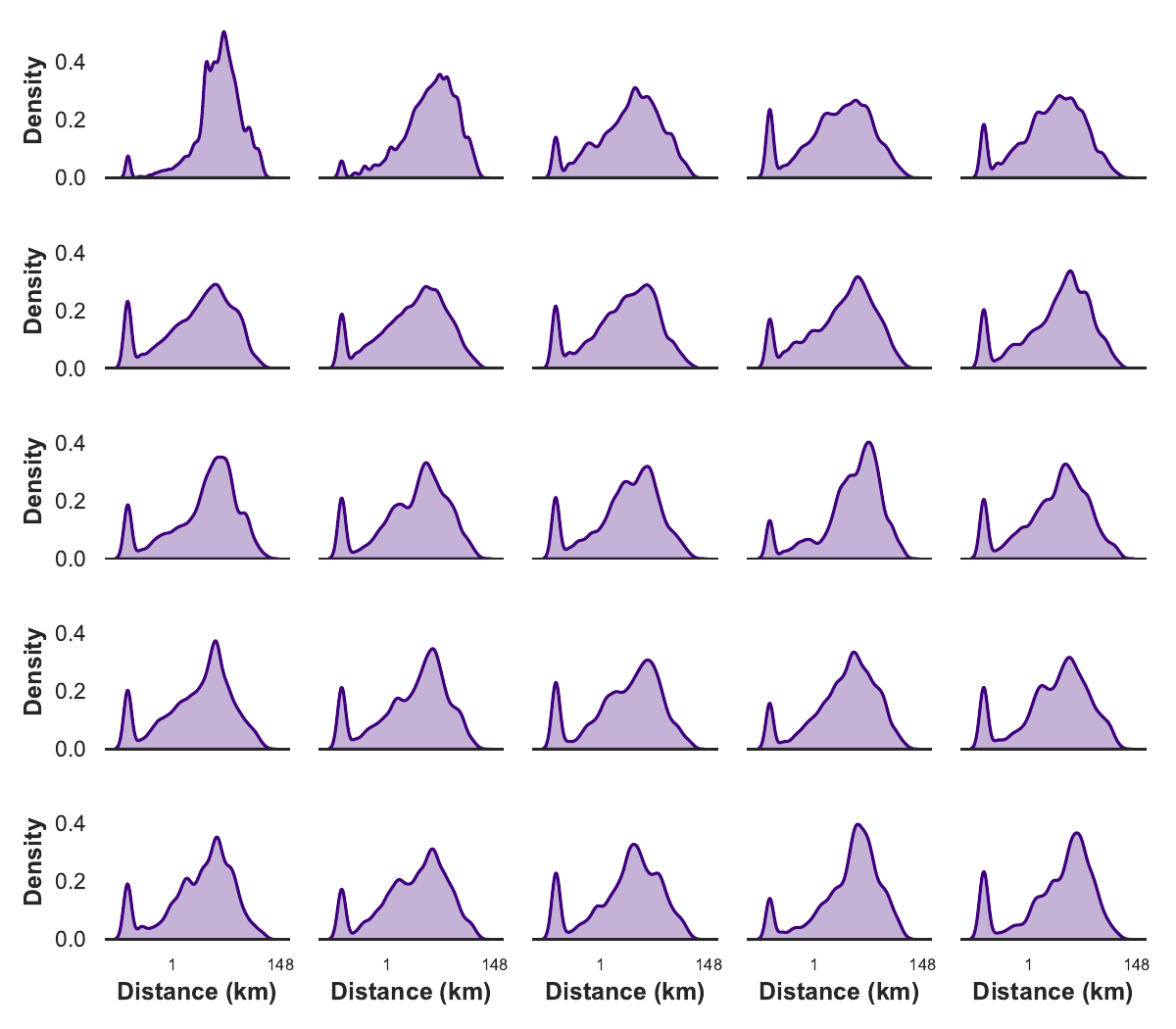}
\caption{\textbf{Trip-length distributions are multimodal.} Kernel density of log trip length for the twenty-five neighbourhoods with the most trips in Paris, with distance shown in kilometres. Most neighbourhoods carry more than one mode, the empirical motivation for modelling each neighbourhood's trips as a mixture of three Gamma components.}
\label{si_multimodal}
\end{figure*}

The mixture model rests on a simple observation: a neighbourhood's trips rarely cluster at a single distance. Supplementary Fig.~\ref{si_multimodal} plots the distribution of trip lengths, on a logarithmic scale, for the twenty-five neighbourhoods with the most trips in Paris. Few are unimodal. Most carry two or three peaks---a near peak for local errands, a far peak for journeys across the city---and it is this structure that the three-component mixture is built to recover. A single Gamma would smooth these peaks into one average distance and discard exactly the information we are after.

\clearpage

\section{Expectation--Maximisation estimation}\label{si_em}

We estimate the three-component Gamma mixture in Equation~(1) of the main text separately for each neighbourhood $i$, using the Expectation--Maximisation (EM) algorithm. Let the parameter vector for neighbourhood $i$ be
\[
\boldsymbol{\theta}_i=\{\pi_{ic},\alpha_{ic},\beta_{ic}\}_{c\in\{p,m,d\}},
\]
where $\pi_{ic}$ is the mixing weight for component $c$, and $\alpha_{ic}$ and \(\beta_{ic}\) are the corresponding Gamma shape and rate parameters.

Given the observed trip lengths \(l_i=\{l_{i1},\dots,l_{iN_i}\}\), the incomplete-data likelihood is
\[
\mathcal{L}(\boldsymbol{\theta}_i\mid l_i)
=\prod_{j=1}^{N_i}\sum_{c\in\{p,m,d\}}
\pi_{ic}\,\Gamma(l_{ij}\mid\alpha_{ic},\beta_{ic}).
\]

We use EM to iteratively estimate the parameters.

\bigskip
\noindent
\textbf{E-step.} Given the current parameter estimates, we compute the posterior probability that trip \(j\) from neighbourhood \(i\) with length \(l_{ij}\) belongs to component \(c\),
\[
\gamma_{ijc}
= \frac{\pi_{ic}\,\Gamma(l_{ij}\mid\alpha_{ic},\beta_{ic})}
       {\sum_{c'}\pi_{ic'}\,\Gamma(l_{ij}\mid\alpha_{ic'},\beta_{ic'})}.
\tag{S1}
\]

\noindent
These posterior probabilities sum to one across components for each trip.

\bigskip
\noindent
\textbf{M-step.} In the M-step, the posterior probabilities \(\gamma_{ijc}\) are treated as fractional component memberships, which are then used to update the mixture weights and component-specific Gamma parameters
\begin{gather}
\pi_{ic}^{\text{new}} =
  \frac{1}{N_i}\sum_{j}\gamma_{ijc},\nonumber\\[4pt]
\bar l_{ic} =
  \frac{\sum_{j}\gamma_{ijc}\,l_{ij}}
       {\sum_{j}\gamma_{ijc}},\qquad
\overline{\ln l}_{ic} =
  \frac{\sum_{j}\gamma_{ijc}\,\ln l_{ij}}
       {\sum_{j}\gamma_{ijc}},\nonumber\\[4pt]
s_{ic} = \ln\bar l_{ic}\;-\;\overline{\ln l}_{ic},\nonumber\\[4pt]
\alpha_{ic}^{\text{new}} =
  h^{-1}(s_{ic}),\quad
  \text{where } h(\alpha)=\psi(\alpha)-\ln\alpha,\nonumber\\[4pt]
\beta_{ic}^{\text{new}} =
  \frac{\alpha_{ic}^{\text{new}}}{\bar l_{ic}}.
\tag{S2}
\end{gather}

\noindent
Here \(N_i\) is the number of trips observed from neighbourhood \(i\), and \(\psi(\cdot)\) denotes the digamma function. The first line updates the component weights using the sample shares weighted with $\gamma_{ijc}$. The remaining lines update the Gamma parameters so that the new component mean and log-moment match the sample weighted with $\gamma_{ijc}$.

\section{Initialisation and convergence}

The EM algorithm is initialised separately for each neighbourhood. To obtain plausible starting values for the three components, we first apply \(k\)-means clustering with \(k=3\) to the log-transformed trip lengths, \(\log l_i\). Log transformation stabilises the wide span of observed distances and improves the separation between short-, intermediate- and long-range trips. The resulting clusters provide an initial partition of trips into proximal, medial and distal groups, from which starting values for \(\pi_{ic}\), \(\alpha_{ic}\) and \(\beta_{ic}\) are computed. These define the initial parameter vector \(\boldsymbol{\theta}^{(0)}_i\).

From these starting values, the EM algorithm alternates between the E-step and M-step until convergence. Convergence is defined as the point at which the increase in log-likelihood between successive iterations falls below \(10^{-6}\). This procedure is repeated independently for each neighbourhood-level trip-length distribution.

After convergence, the estimated components are ordered by increasing mean,
\[
\hat{\mu}_{ic}=\frac{\alpha_{ic}}{\beta_{ic}},
\]
and labelled proximal (\(p\)), medial (\(m\)) and distal (\(d\)). This ordering ensures a consistent interpretation of the components across neighbourhoods and cities.

\clearpage

\section{Selection of the number of components}\label{si_bic}

We selected the number of mixture components by comparing Bayesian Information Criterion (BIC) values across alternative specifications. For each neighbourhood, we estimated finite mixtures with different numbers of components and computed
\[
\mathrm{BIC} = -2\log \hat{\mathcal{L}} + k \log N_i,
\]
where \(\hat{\mathcal{L}}\) is the maximised likelihood, \(k\) is the number of free parameters in the model, and \(N_i\) is the number of trips observed from neighbourhood \(i\). Lower BIC values indicate a better trade-off between goodness of fit and model complexity.

Across all neighbourhoods, the three-component specification minimised BIC in approximately \(90\%\) of origins. We therefore retain three components as the preferred specification, since it provides a parsimonious and interpretable decomposition of mobility into proximal, medial and distal scales. We report the comparison across specifications in Supplementary Fig.~\ref{si_model_selection}.

\begin{figure*}[!ht]
\centering
\includegraphics[width=0.9\textwidth]{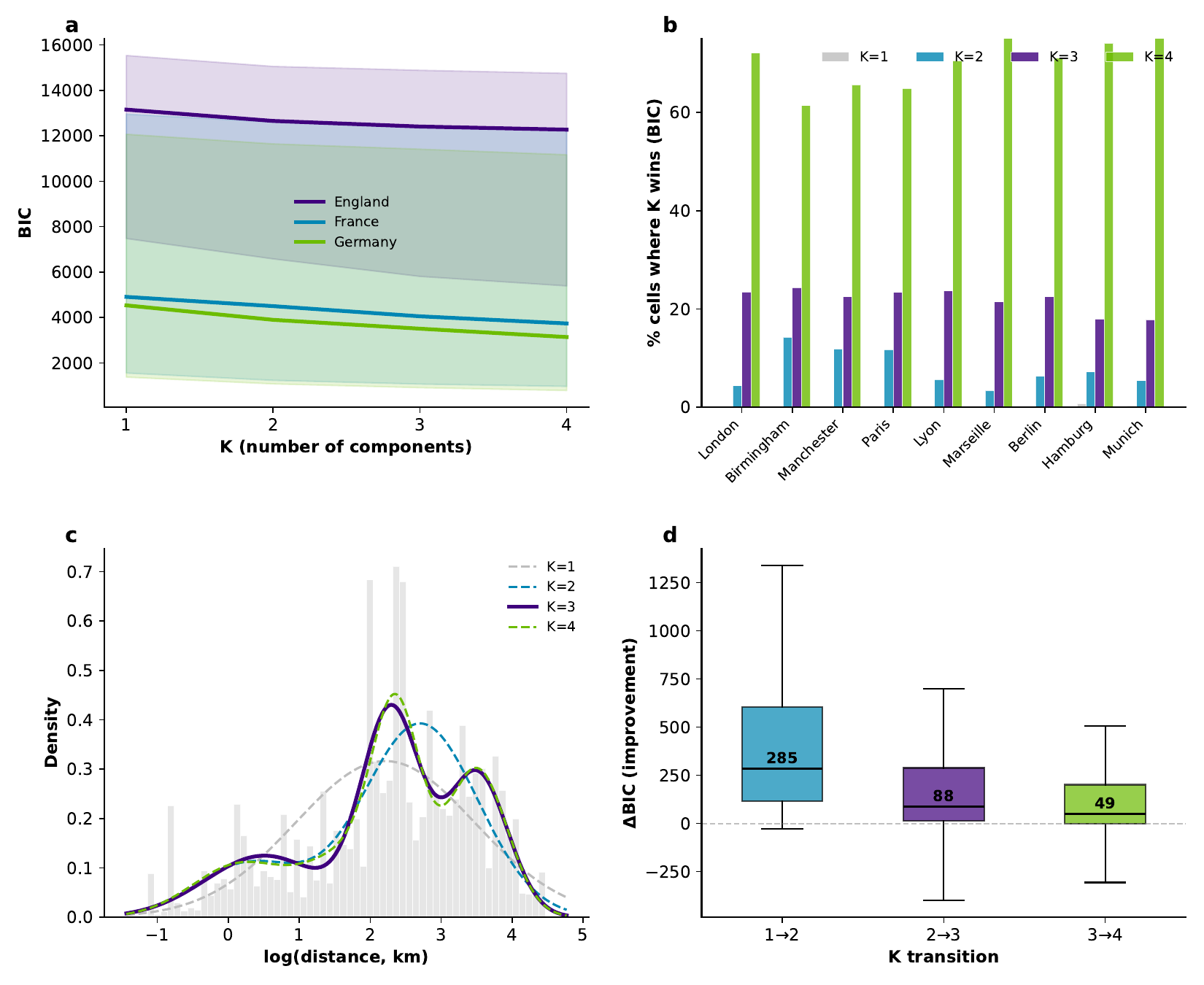}
\caption{\textbf{Three components win across neighbourhoods.} We fit finite mixtures of one to four components to log trip lengths. \textbf{a} BIC falls steeply to three components and flattens thereafter, with a line per country and the interquartile range across neighbourhoods shaded. \textbf{b} Three components minimise BIC in roughly nine neighbourhoods in ten, city by city. \textbf{c--d} Fitted densities laid over the observed log-distance histogram trace the near and far modes that fewer components miss.}
\label{si_model_selection}
\end{figure*}

\clearpage

\section{Quantifying socioeconomic mixing}\label{si_mixing}

To examine how different mobility scales contribute to experienced socioeconomic mixing, we classify each trip according to the component with the highest posterior probability \(\gamma_{ijc}\), given by Eq.~(S1). This yields a partition of trips into proximal, medial and distal mobility scales.

We assign both home neighbourhoods and destination hexagons to one of five socioeconomic quintiles using harmonised small-area indicators within each country, described in Supplementary Section~\ref{si_data}. In England, these quintiles are based on area-level deprivation; in France, on income; and in Germany, on rent. Quintiles are constructed within country to capture relative socioeconomic position in each national context.

For each neighbourhood \(i\) and mobility component \(c\), we compute the distribution of destination quintiles reached by its residents,
\[
p_{igc}
=
\frac{
\sum_j \mathbf{1}\{Q_{ij}=g,\; \hat c_{ij}=c\}
}{
\sum_j \mathbf{1}\{\hat c_{ij}=c\}
},
\tag{S4}
\]
where \(Q_{ij}\) denotes the socioeconomic quintile of the destination reached on trip \(j\) from neighbourhood \(i\), \(\hat c_{ij}\) is the component to which trip \(j\) is assigned, and \(\mathbf{1}\{\cdot\}\) is the indicator function. The vector
\[
\mathbf{p}_{ic} = (p_{i1c},\dots,p_{i5c})
\]
therefore summarises the socioeconomic composition of destinations reached by residents of neighbourhood \(i\) through mobility scale \(c\).

Our main measure of experienced socioeconomic mixing is the Hill number of order \(1\),
\[
\mathrm{Hill}_{1,ic}
=
\exp\!\left(
-\sum_{g=1}^{5} p_{igc}\ln p_{igc}
\right).
\tag{S5}
\]
This quantity can be interpreted as the effective number of socioeconomic groups encountered through mobility scale \(c\). It increases both with the number of groups encountered and with the evenness of exposure across them, so that higher values indicate broader and more balanced socioeconomic mixing.

For robustness, we also compute the Hill numbers of order \(0\) and \(2\),
\begin{align}
\mathrm{Hill}_{0,ic}
&=
\left|
\left\{
g : p_{igc}>0
\right\}
\right|,\nonumber\\[4pt]
\mathrm{Hill}_{2,ic}
&=
\left(
\sum_{g=1}^{5} p_{igc}^{\,2}
\right)^{-1}.
\tag{S6}
\end{align}
We report these alternative orders in Supplementary Fig.~\ref{si_diversity}.

At the city level, we additionally summarise mixing patterns using \(5\times5\) interaction matrices for each mobility scale,
\[
P_c(g' \mid g)
=
\frac{
\sum_i \sum_j \mathbf{1}\{Q_i=g,\;Q_{ij}=g',\;\hat c_{ij}=c\}
}{
\sum_i \sum_j \mathbf{1}\{Q_i=g,\;\hat c_{ij}=c\}
},
\tag{S7}
\]
where \(Q_i\) is the socioeconomic quintile of home neighbourhood \(i\), and \(Q_{ij}\) is the quintile of the destination reached on trip \(j\). Each row of \(P_c(g' \mid g)\) therefore gives the probability that residents of home quintile \(g\) reach destinations in quintile \(g'\) through mobility scale \(c\). We give these matrices city by city and scale by scale in Supplementary Section~\ref{si_matrices}.

\clearpage

\section{Mobility and relative centrality}\label{si_centrality}

\begin{figure*}[!ht]
\centering
\includegraphics[width=0.95\textwidth]{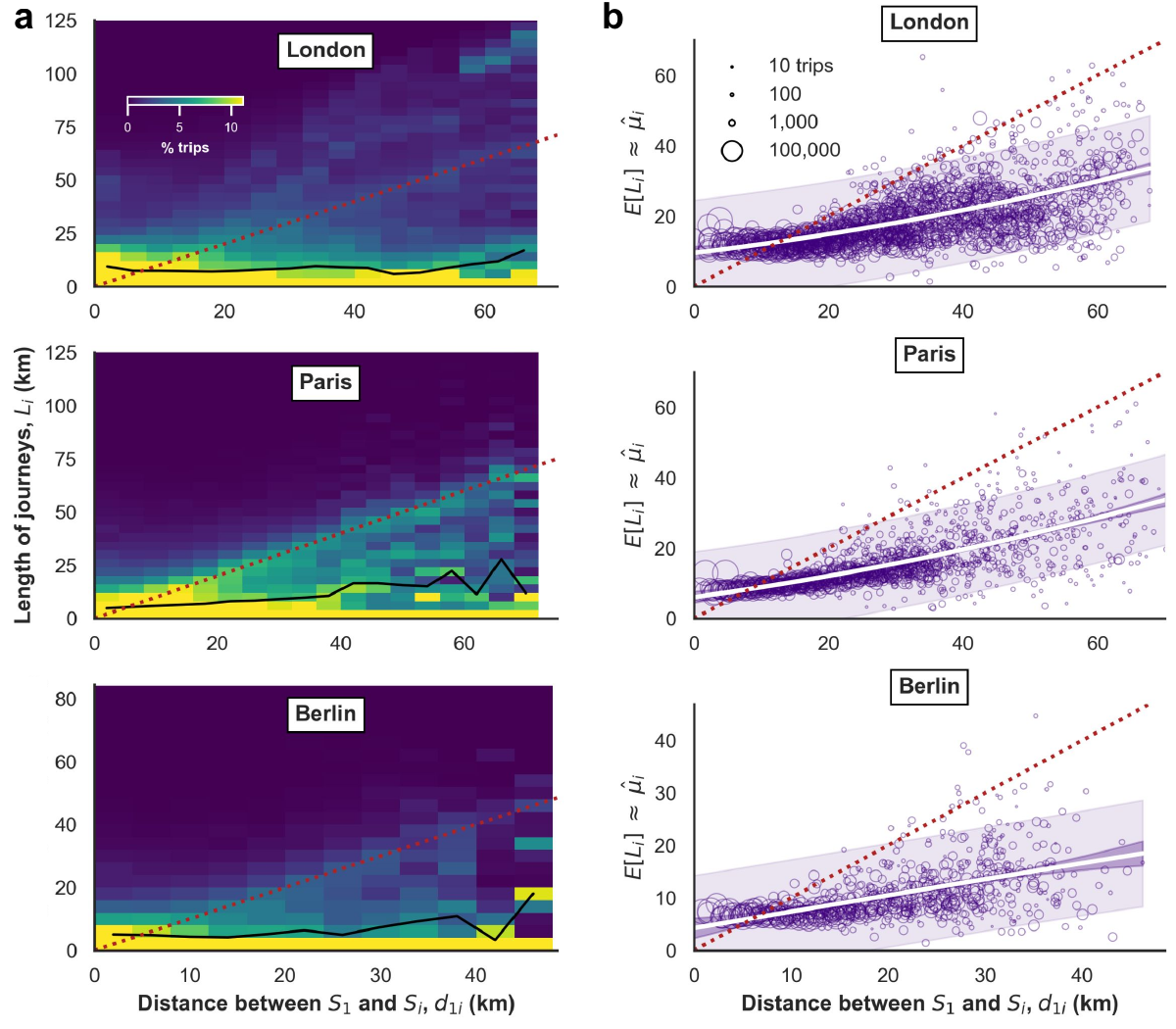}
\caption{\textbf{Trips fall far short of a single-centre city.} \textbf{a} Binned against distance from the primary centre $d_{1i}$, the distribution of journey lengths $L_i$ sits well below the 1:1 line a single-centre city would follow (dashed), for London, Paris, and Berlin; the solid line marks the median. \textbf{b} Each neighbourhood's mean trip length $\mathbb{E}[L_i]\approx\hat\mu_i$ rises with distance from the centre but stays beneath that line, traced by a quadratic fit with confidence and prediction bands; marker size scales with the number of trips. The data sit well below the 1:1 line throughout, so trips are far shorter than a monocentric city would require.}
\label{si_centrality_fig}
\end{figure*}

If a city turned on a single centre, every trip would run between a neighbourhood and that centre, and a trip's length would simply be the neighbourhood's distance from it. Supplementary Fig.~\ref{si_centrality_fig} sets that monocentric expectation---the dashed 1:1 line---against what residents actually do, for London, Paris, and Berlin. They fall well short of it. The bulk of trips, and the median in every city (the solid line in panel \textbf{a}), sit far below the 1:1 line: most journeys are much shorter than a trip to the centre would be, and they grow with distance from the centre only gently rather than one-for-one. Panel \textbf{b} tells the same story for the mean trip length of each neighbourhood, $\mathbb{E}[L_i]\approx\hat\mu_i$, which rises with centrality but stays beneath the single-centre line throughout. A single component cannot carry this: collapsing a neighbourhood's trips to one characteristic distance would place it on the 1:1 line, exactly where the data are not. The gap is what the three-scale mixture is built to describe.

\clearpage

\section{Variation in mobility scales across European cities}

\begin{figure*}[!ht]
\centering
\includegraphics[width=1\textwidth]{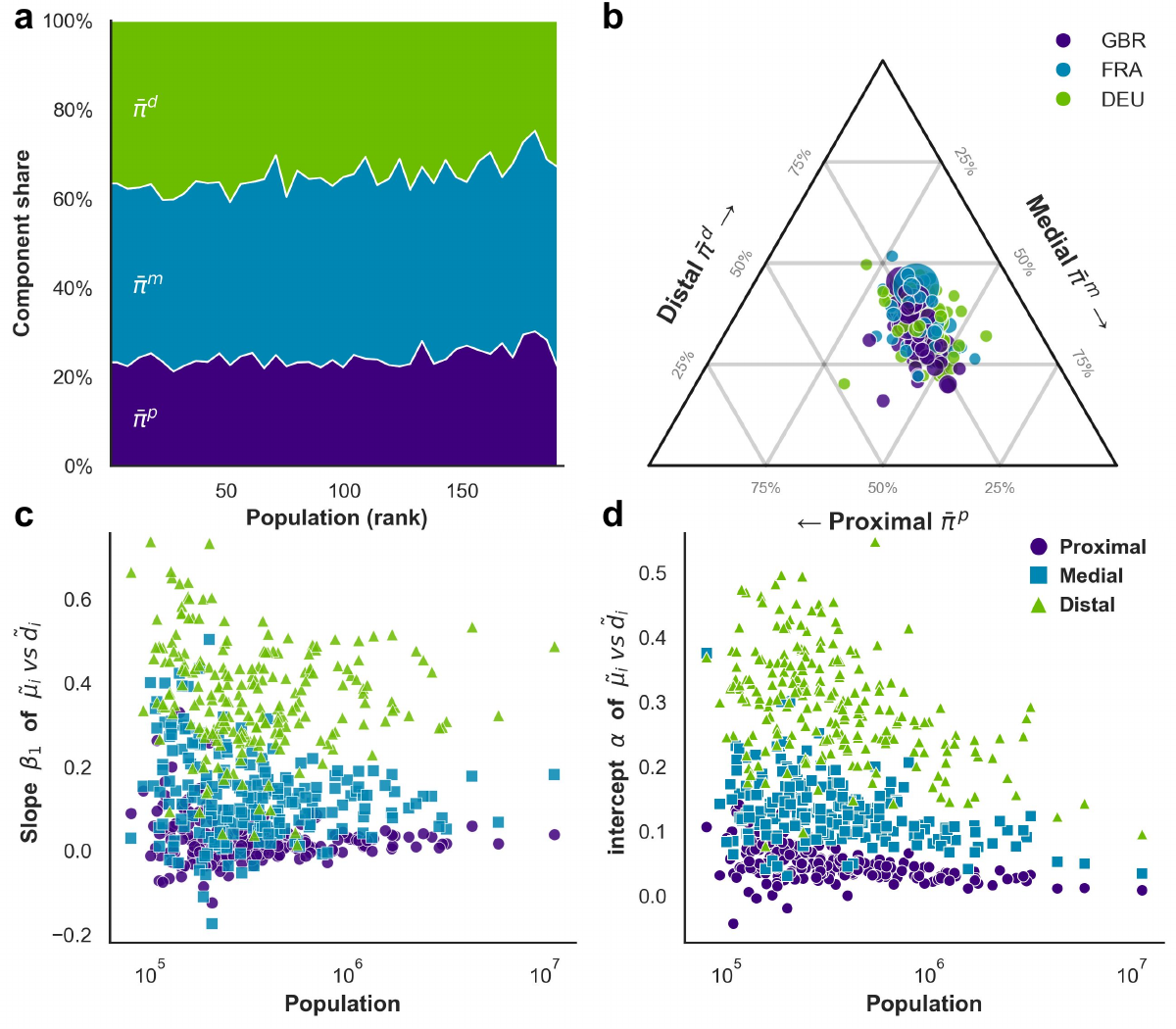}
\caption{\textbf{A stable mix of scales, shortening with city size.} \textbf{a} Average component weights, ordered by city population rank, hold their relative shares of proximal, medial, and distal trips across cities. \textbf{b} A ternary plot places each city by its balance of the three components, coloured by country; the cities cluster rather than scatter. \textbf{c} The slope \(b_c\) from Equation~(2) of the main text declines with city population, \textbf{d} as does the intercept \(a_c\), most steeply for the distal scale---so as cities grow their long trips shorten and their structure leans less on a single nucleus.}
\label{si_variation}
\end{figure*}

We use Supplementary Fig.~\ref{si_variation} to examine how the fitted mobility scales vary across cities, beyond the patterns shown in Fig.~4 of the main text. In particular, we ask whether the relative weights of the proximal, medial and distal components vary systematically across cities, and whether the relationship between mobility scale and distance from the nucleus changes with city size.

Panel \textbf{a} shows the average component weights across cities ordered by population rank. Panel \textbf{b} offers an alternative visualisation as a ternary plot, where colour indicates country and position in the triangle indicates the relative contribution of the proximal, medial and distal components to the average mobility profile of each city. Panels \textbf{c} and \textbf{d} then relate the estimated city-level parameters from Equation~(2) of the main text to city population size, showing how the slope \(b_c\) and intercept \(a_c\) vary across cities.

Two findings stand out. First, from panels \textbf{a} and \textbf{b}, the relative weights of the proximal, medial and distal components are remarkably stable across cities and countries. The proximal component is typically the smallest share of trips, whereas the medial and distal components account for larger shares, indicating that most daily needs are met beyond the immediate residential environment. Second, there is a city-size effect on the scale of the distal component. Both intercepts and slopes in panels \textbf{c} and \textbf{d} show a tendency to decline with city size for the distal scale. This implies that long trips tend to become shorter for residents of larger cities, suggesting that a more polycentric organisation emerges with urban growth. While the three mobility scales are a widespread feature of everyday travel, larger cities are modestly less dependent on a single dominant nucleus and more consistent with a polycentric organisation of daily activity.

\clearpage

\section{Mobility scales in smaller cities}\label{si_smaller_cities}

\begin{figure*}[!ht]
\centering
\includegraphics[width=0.95\textwidth]{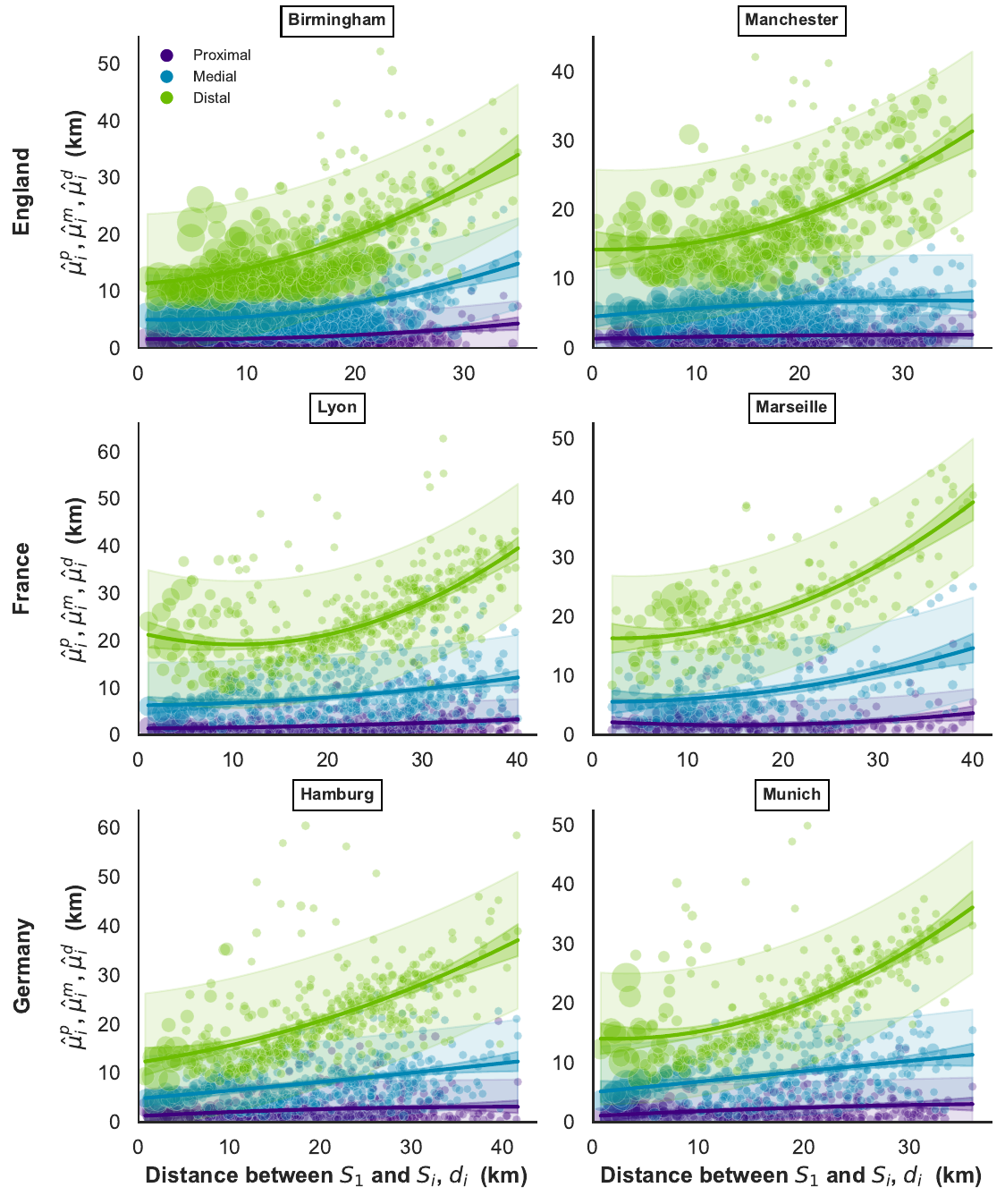}
\caption{\textbf{The three-scale gradient holds in smaller cities.} The mean trip length of each neighbourhood at the proximal, medial, and distal scale against its distance from the city centre, for the second- and third-largest city in each country: Birmingham and Manchester (England), Lyon and Marseille (France), Hamburg and Munich (Germany). Each panel carries a quadratic fit per scale with its confidence and prediction bands, and marker size scales with the number of trips. As in the three capitals, the proximal mean stays flat, the distal mean rises steeply toward the periphery, and the medial mean falls between them.}
\label{si_component_scatter_cities}
\end{figure*}

We repeat the component scatter of Fig.~3 of the main text---each scale's mean trip length against a neighbourhood's distance from the centre---for the second- and third-largest city in each country: Birmingham and Manchester, Lyon and Marseille, Hamburg and Munich. Supplementary Fig.~\ref{si_component_scatter_cities} sets them side by side, and the same gradient returns in every one. The proximal mean stays flat across the city, local life running at a common short scale wherever a neighbourhood sits; the distal mean climbs steeply toward the periphery, the signature of journeys that must cross the city when home lies far from the centre; and the medial mean rises gently between the two. The pattern is, if anything, sharper in these smaller cities than in the capitals, so the three-scale structure---and the core--periphery gradient it carries---is a general feature of how these cities move rather than a peculiarity of London, Paris, or Berlin.

\clearpage

\section{Maps of component weights and model fit}\label{si_maps}

\begin{figure*}[!ht]
\centering
\includegraphics[width=0.95\textwidth]{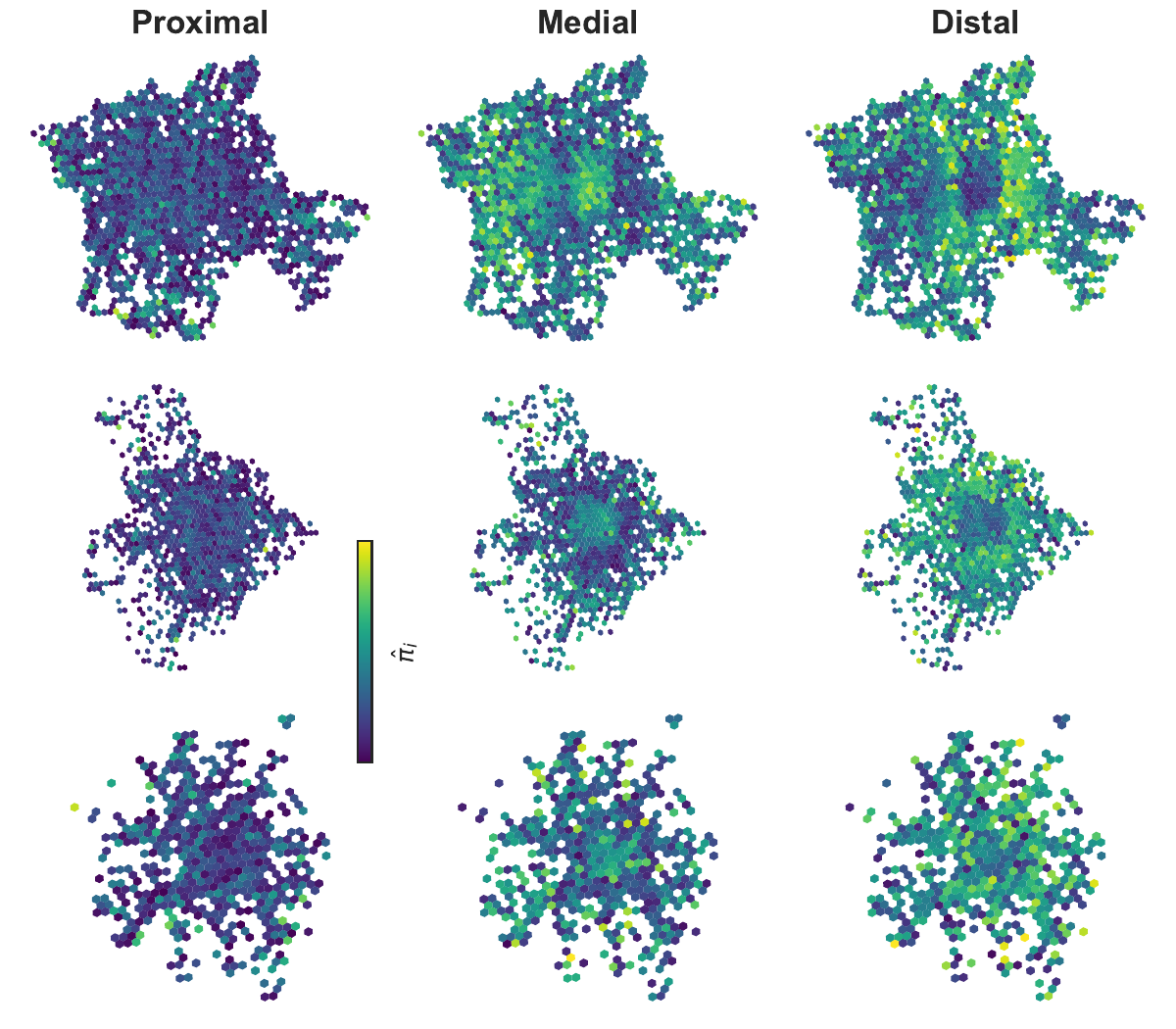}
\caption{\textbf{Proximal and medial trips concentrate toward the centre.} The share of trips at the proximal, medial, and distal scale for each neighbourhood of London, Paris, and Berlin. Proximal and medial weights concentrate toward the centre; the distal weight stays low there, where the fitted distal component points outward rather than toward the core.}
\label{si_weights}
\end{figure*}

\begin{figure*}[!ht]
\centering
\includegraphics[width=0.95\textwidth]{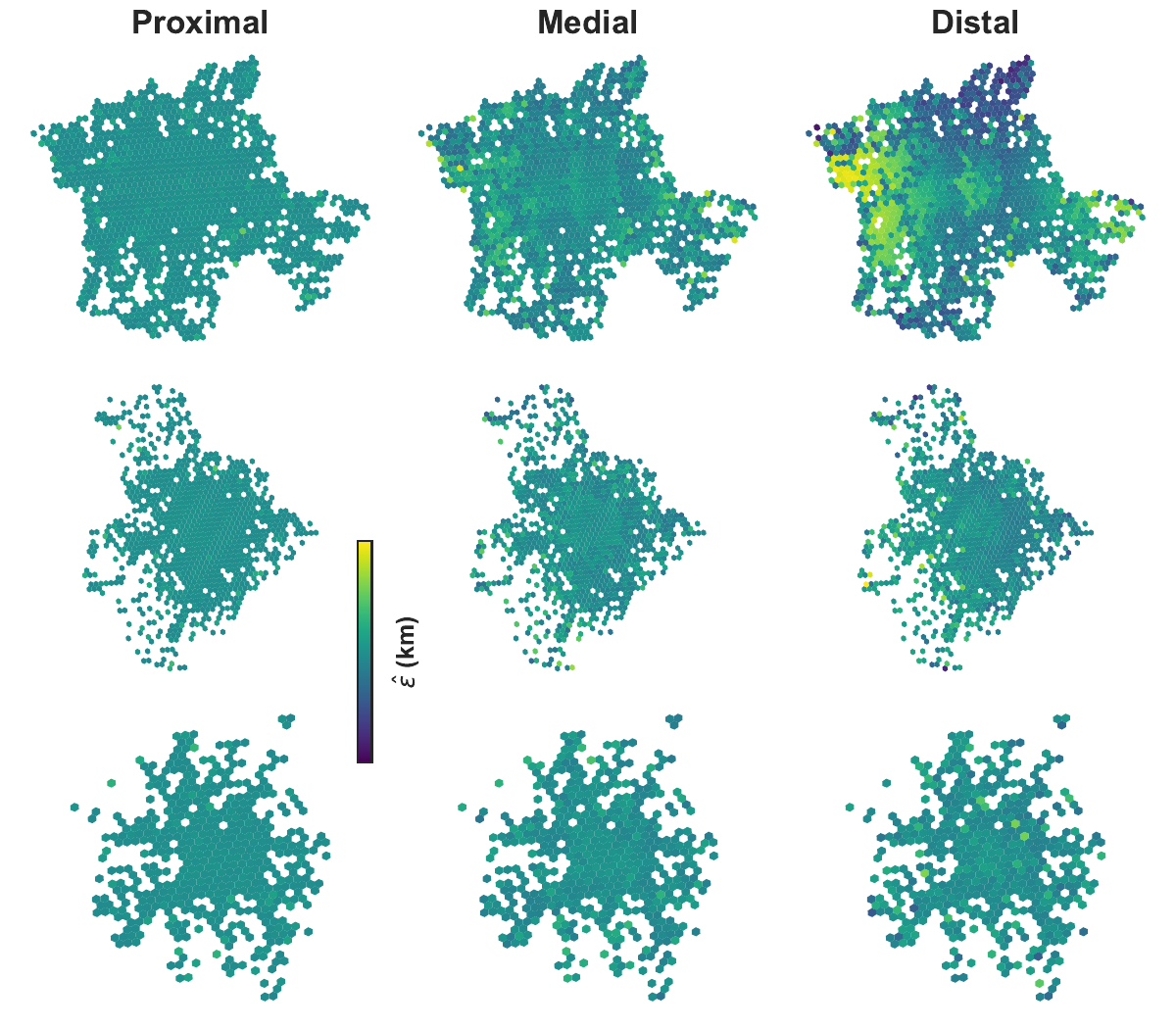}
\caption{\textbf{The mixture fits worst at the periphery.} Residuals from the three-component mixture for London, Paris, and Berlin. Fit is poorest at the periphery, where trips spread across more characteristic distances than three components separate cleanly---a pattern that itself signals a city operating at more than one scale.}
\label{si_residuals}
\end{figure*}

This section maps two diagnostics across London, Paris, and Berlin. Supplementary Fig.~\ref{si_weights} maps the component weights---the share of a neighbourhood's trips that each scale carries. The proximal and medial weights are highest toward the centre, where local and intermediate trips meet most daily needs, while the distal weight is low there: the model still fits a distal component to central neighbourhoods, but it carries little of their travel and points outward rather than in. Supplementary Fig.~\ref{si_residuals} maps the model residuals. Fit is poorest at the periphery, where a neighbourhood's trips spread across more characteristic distances than three components can cleanly separate---itself a sign of a city operating at several scales at once.

\clearpage

\section{Interaction matrices by city and component}\label{si_matrices}

\begin{figure}[p]
\centering
\includegraphics[width=0.92\textwidth]{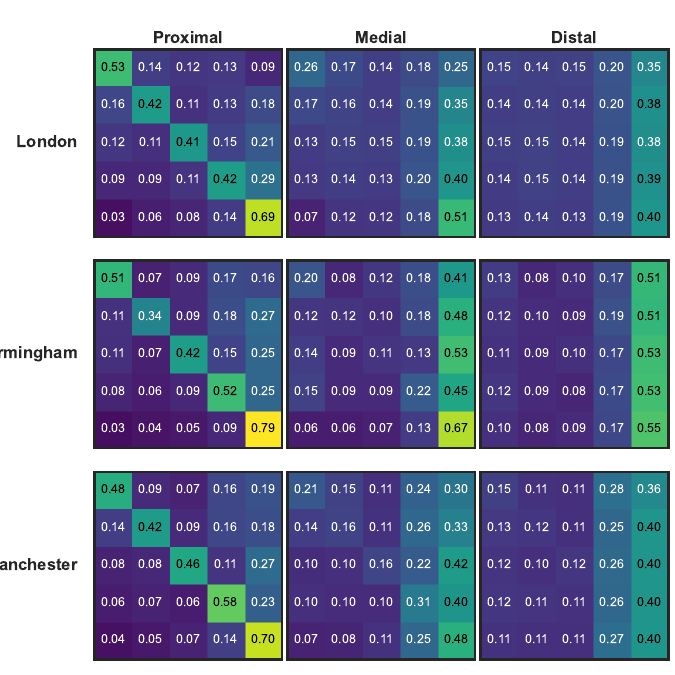}
\caption{\textbf{Mixing runs upward across the English cities.} Rows are London, Birmingham, and Manchester; columns are the proximal, medial, and distal scales. Each $5\times5$ matrix gives the probability $P(j\mid i)$ that a resident of home quintile $i$ (row) reaches a destination in quintile $j$ (column), from the least advantaged quintile (top left) to the most. A common colour scale spans all panels. Mass above the diagonal---lower-status residents reaching higher-status destinations---exceeds the mass below it, the upward bias discussed in the text.}
\label{si_matrices_eng}
\end{figure}

\begin{figure}[p]
\centering
\includegraphics[width=0.92\textwidth]{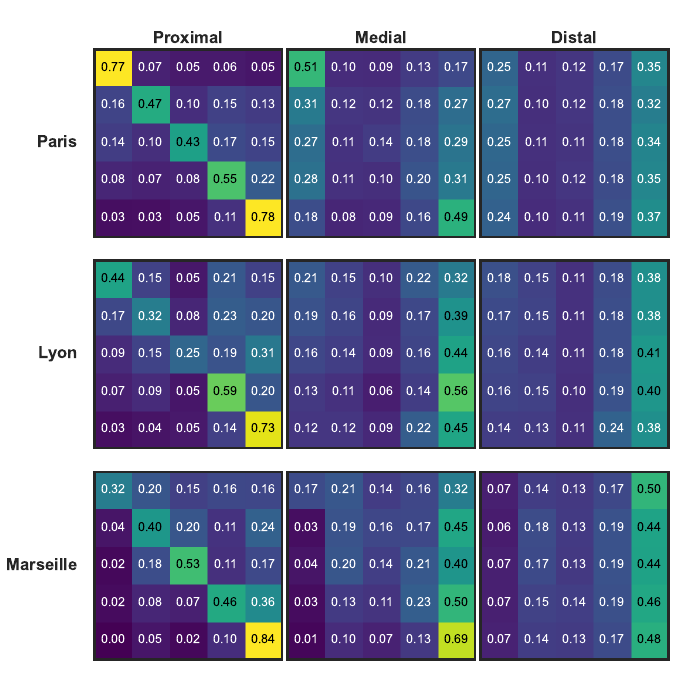}
\caption{\textbf{Mixing runs upward across the French cities.} Rows are Paris, Lyon, and Marseille; columns are the proximal, medial, and distal scales. Each $5\times5$ matrix gives the probability $P(j\mid i)$ that a resident of home quintile $i$ reaches a destination in quintile $j$, from the least advantaged quintile to the most, on a colour scale shared with the other countries. The upward bias above the diagonal recurs here as in England.}
\label{si_matrices_fra}
\end{figure}

\begin{figure}[p]
\centering
\includegraphics[width=0.92\textwidth]{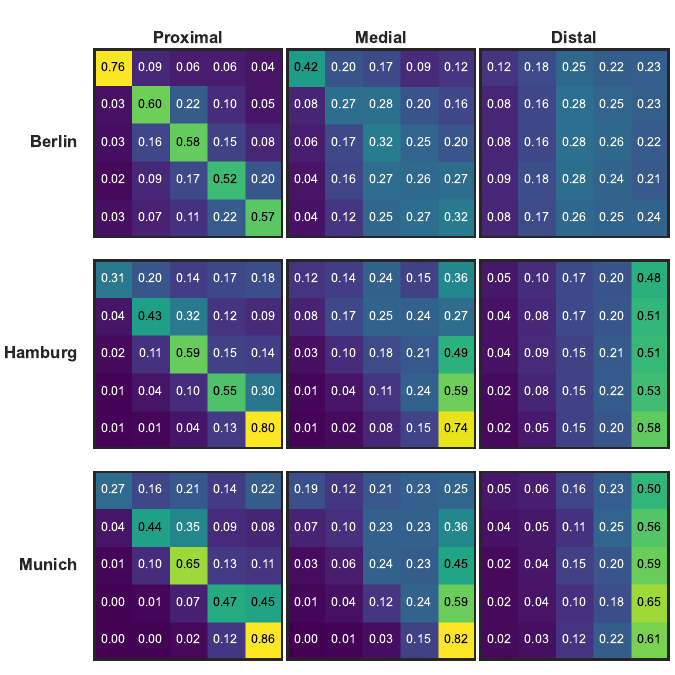}
\caption{\textbf{Mixing runs upward across the German cities.} Rows are Berlin, Hamburg, and Munich; columns are the proximal, medial, and distal scales. Each $5\times5$ matrix gives the probability $P(j\mid i)$ that a resident of home quintile $i$ reaches a destination in quintile $j$, from the least advantaged quintile to the most, on a colour scale shared with the other countries. The same upward bias holds, completing the pattern across all three countries.}
\label{si_matrices_ger}
\end{figure}

The interaction matrices in the main text pool the nine cities into a single panel. Here we break them out city by city and scale by scale, both to show how the pattern holds across very different urban and social structures and to make its one consistent asymmetry visible. For each city we give three matrices---proximal, medial, and distal. Each gives the probability $P(j\mid i)$ that a resident of home quintile $i$ reaches a destination in quintile $j$, with rows running from the least advantaged quintile to the most and a common colour scale across every panel. Supplementary Figs.~\ref{si_matrices_eng}, \ref{si_matrices_fra}, and~\ref{si_matrices_ger} take the three cities of England, France, and Germany in turn.

Two features recur in every city and at every scale. The first is homophily: the mass sits on the diagonal, and residents most often reach destinations of their own standing. The second is an upward bias off that diagonal. The mass above the diagonal---lower-status residents reaching higher-status destinations---is consistently heavier than the mass below it, where higher-status residents reach lower-status ones. Residents of poorer neighbourhoods travel into richer parts of the city far more than the reverse, a pull that tracks the gradient of rent placing wealthier residents, and the amenities they support, near the core.

The bias is not uniform, and that is the point of breaking the matrices apart. It strengthens from the proximal to the distal scale: short trips stay close to home in every sense, reaching destinations of similar standing, while the longest trips carry residents furthest up the socioeconomic gradient. It also varies in degree from city to city---some show a near-even exchange at the proximal scale and a pronounced upward pull only at the distal, others carry the asymmetry across all three scales---but its direction does not reverse. In none of the nine cities do higher-status residents reach lower-status destinations more often than the reverse. The upward bias is, on this evidence, a robust feature of how European cities mix.


\clearpage

\clearpage

\clearpage

\section{Mixing across the centre--periphery gradient}\label{si_distance_decay}

\begin{figure*}[!ht]
\centering
\includegraphics[width=0.95\textwidth]{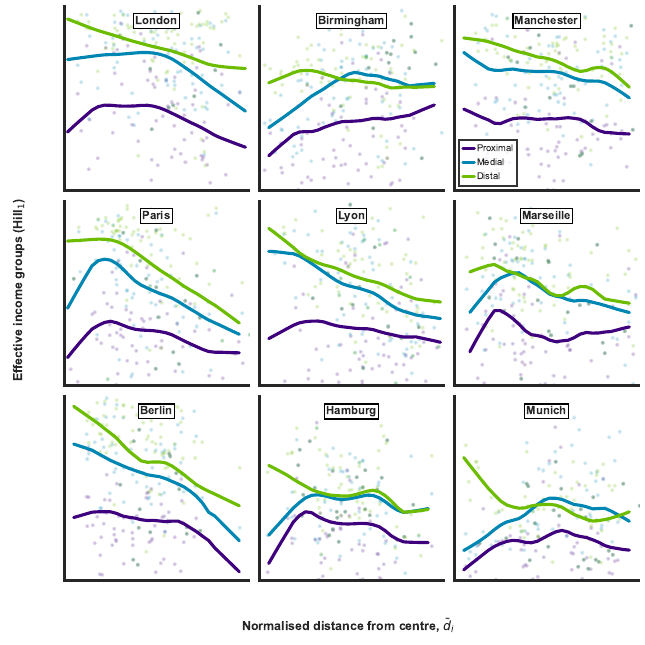}
\caption{\textbf{Mixing falls toward the periphery at every scale.} The effective number of income groups reached (Hill$_1$) against normalised distance from the city centre $\tilde d_i$, for the proximal, medial, and distal scales, in each of the nine cities. Curves are smoothed fits over neighbourhoods. The distal scale stays most diverse across the gradient, and mixing declines toward the periphery at every scale.}
\label{si_distance_decay_fig}
\end{figure*}

Mixing is not only higher on longer trips; it also falls as residents live farther from the centre, and it does so at every scale. Supplementary Fig.~\ref{si_distance_decay_fig} plots the effective number of income groups reached (Hill$_1$) against a neighbourhood's normalised distance from the centre, separately for the proximal, medial, and distal scales, in each of the nine cities. Two features hold across cities: the distal curve sits above the medial and the medial above the proximal, so the ordering of the scales by mixing is stable; and all three slope downward, so peripheral residents encounter a narrower range of groups than central ones, whichever scale they travel at.

\clearpage

\section{Counterfactual simulations}\label{si_counterfactual}

To distinguish the role of destination choice from the role of longer-distance travel in generating socioeconomic mixing, we construct three counterfactual mobility regimes.

First, we define a null mobility regime that preserves each neighbourhood's total trip volume and fitted trip-length distribution while randomising destinations. This provides a benchmark for the level of mixing that would arise if residents travelled just as much and just as far as observed, but to different places.

Second, we construct two modified regimes that remove the distal mobility component while preserving each neighbourhood's total trip volume. For each neighbourhood \(i\), let the fitted mixture weights be \((\pi_{ip},\pi_{im},\pi_{id})\). In the first case, the distal mass is reassigned entirely to the medial component,
\[
(\pi_{ip}',\pi_{im}',\pi_{id}')
=
(\pi_{ip},\;\pi_{im}+\pi_{id},\;0).
\tag{S8}
\]
In the second case, the distal mass is reassigned entirely to the proximal component,
\[
(\pi_{ip}',\pi_{im}',\pi_{id}')
=
(\pi_{ip}+\pi_{id},\;\pi_{im},\;0).
\tag{S9}
\]

For each counterfactual regime, we simulate trip lengths while preserving the total number of trips originating from each neighbourhood. Conditional on simulated trip length, destinations are sampled from the empirical set of reachable hexagons at the corresponding distance. Under the null regime, this destination sampling randomises interactions while preserving the observed distance structure. Under the distal-removal regimes, it allows us to isolate the contribution of longer-range mobility to socioeconomic mixing while keeping shorter-distance travel patterns intact.

\begin{figure*}[!ht]
\centering
\includegraphics[width=0.7\textwidth]{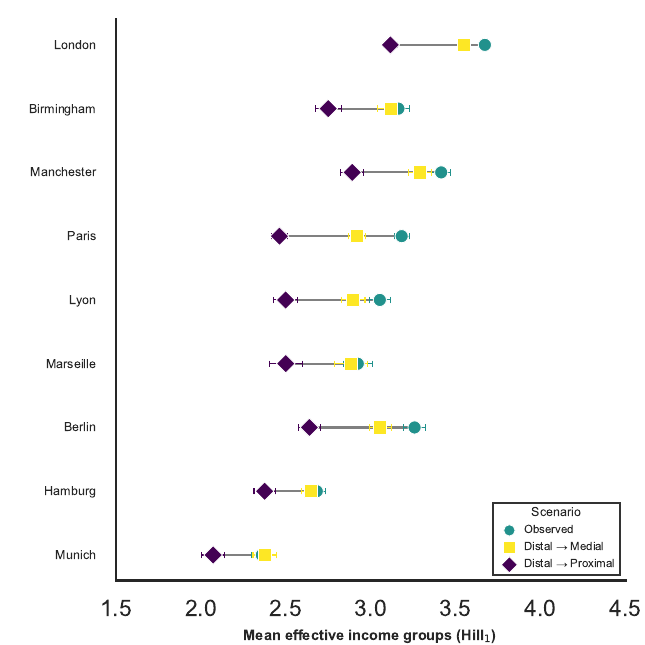}
\caption{\textbf{Removing the longest trips lowers mixing in every city.} The mean effective number of income groups reached (Hill$_1$) for each of the nine cities under the observed mobility (circle), under reassigning the distal component to the medial scale (square), and under reassigning it to the proximal scale (diamond). Error bars span the simulation runs. Removing the longest trips lowers mixing in every city, and reassigning them to the shortest scale costs the most.}
\label{si_counterfactual_fig}
\end{figure*}

\begin{figure*}[!ht]
\centering
\includegraphics[width=0.95\textwidth]{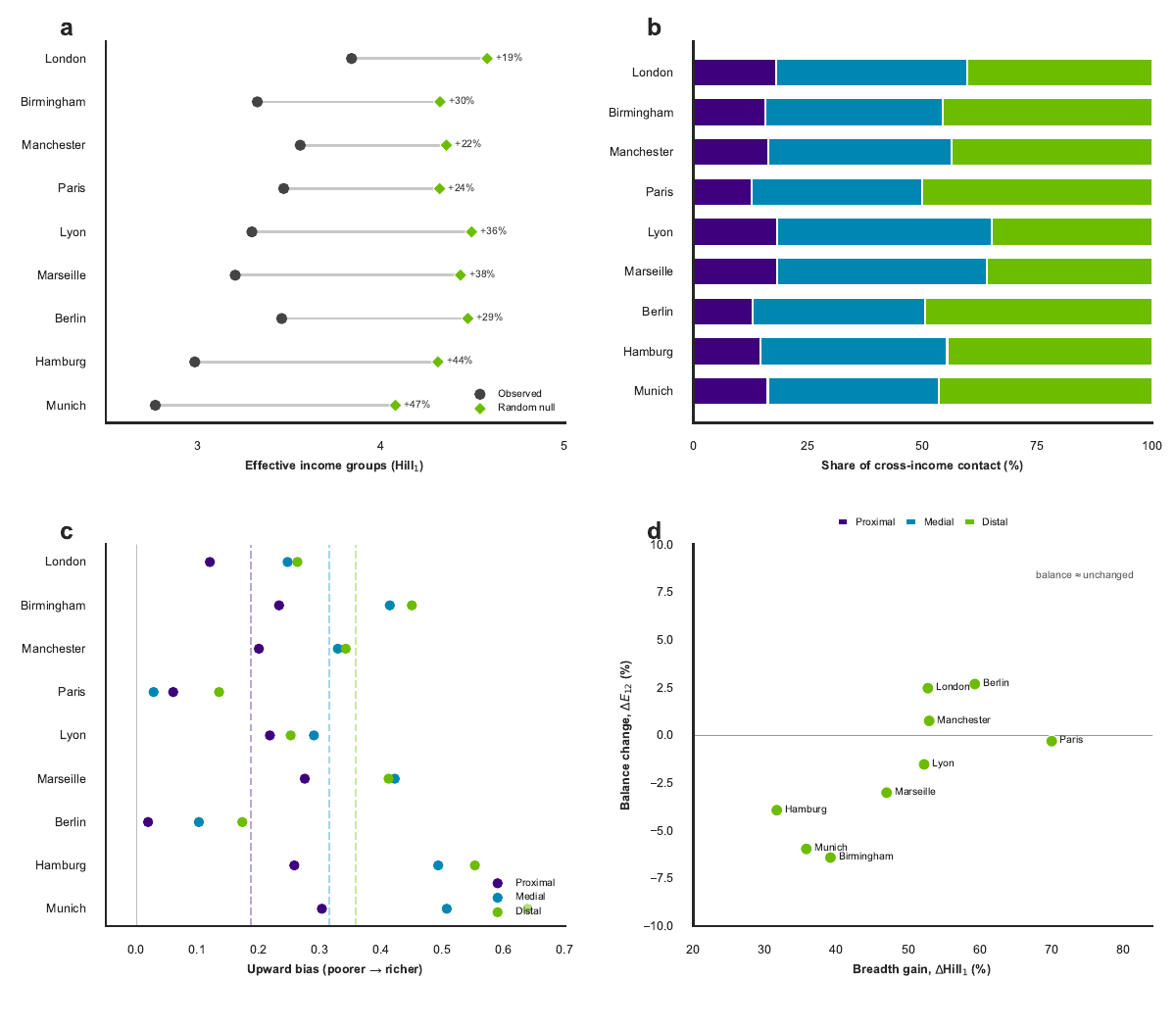}
\caption{\textbf{Long trips drive cross-income mixing.} \textbf{a} Observed mixing ($\mathrm{Hill}_1$) against a random-destination null for each city, with the percentage gap labelled; the null exceeds the observed everywhere, so destination choice is selective. \textbf{b} Share of cross-income contact carried by the proximal, medial, and distal scales; distal trips carry a disproportionate share. \textbf{c} Upward bias (poorer-to-richer contact) by scale and city, with the pooled component means dashed; the bias is strongest at the distal scale. \textbf{d} Breadth against balance: the gain in the number of groups encountered ($\mathrm{Hill}_1$) plotted against the change in evenness ($E_{12}$); distal mobility raises breadth while balance stays roughly unchanged.}
\label{si_mixing_2x2}
\end{figure*}

We then recompute the destination socioeconomic distributions \(p_{igc}\), the corresponding Hill numbers, and the city-level interaction matrices under each simulated regime. To summarise the effect of each counterfactual, we report the percentage change in city-level \(\mathrm{Hill}_1\) relative to the observed mobility regime,
\[
\Delta \mathrm{Hill}_1
=
100\times
\frac{
\mathrm{Hill}_1^{\mathrm{cf}}-\mathrm{Hill}_1^{\mathrm{obs}}
}{
\mathrm{Hill}_1^{\mathrm{obs}}
}.
\tag{S10}
\]
Negative values indicate that the counterfactual regime reduces socioeconomic mixing relative to the observed mobility structure. Supplementary Fig.~\ref{si_counterfactual_fig} shows the result city by city: removing the distal scale lowers experienced mixing in every city, and reassigning its trips to the proximal scale costs more than reassigning them to the medial.

\clearpage

\section{Long trips and cross-income mixing}

This figure gathers the evidence that the longest trips do the most to mix incomes, across the nine cities. Panel \textbf{a} sets observed mixing against a random-destination null: in every city residents mix \emph{less} than chance would allow---between 19\% and 47\% less---so where people go is socially selective, not merely how far they travel. Panel \textbf{b} shows where the cross-income contact comes from: distal trips carry a disproportionate share in every city, around 40\% of all cross-group encounters despite being a smaller part of travel. Panel \textbf{c} adds the asymmetry: at each scale, and most at the distal, contact runs upward, with residents of poorer neighbourhoods reaching richer destinations more than the reverse. Panel \textbf{d} draws the limit: moving outward through the scales raises the breadth of groups encountered ($\mathrm{Hill}_1$) but leaves their balance ($E_{12}$) almost unchanged, so distal mobility widens exposure without evening it.

\clearpage

\section{Diversity and evenness of exposure across socioeconomic groups}

We distinguish whether different mobility scales increase the diversity of socioeconomic exposure or make that exposure more evenly distributed across groups. To this end, we compare Hill$_1$ and Hill$_2$. Hill$_1$ captures the effective number of socioeconomic groups encountered, whereas Hill$_2$ gives greater weight to dominant groups and is therefore more sensitive to concentration in exposure. To compare the two, we consider the evenness ratio
\[
E_{12}=\frac{\mathrm{Hill}_2}{\mathrm{Hill}_1},
\]
where values closer to 1 indicate that exposure is distributed more evenly across the groups encountered.

Across all nine cities, distal trips increase both Hill$_1$ and Hill$_2$ relative to proximal trips, indicating that longer journeys expand socioeconomic exposure in ways that are not driven only by rare contacts. However, the evenness ratio \(E_{12}\) varies little across mobility components. The main advantage of distal mobility is therefore diversity of exposure rather than evenness of exposure across socioeconomic groups. These patterns are shown in Supplementary Fig.~\ref{si_diversity}.

In substantive terms, distal mobility expands the range of people residents encounter in daily life, but it does not overcome the tendency for exposure to remain concentrated within a smaller subset of socioeconomic groups. Much of the gain in diversity comes from moving beyond the local environment, with medial trips recovering a substantial share of this exposure and distal trips providing an additional increment.

\begin{figure*}[!ht]
\centering
\includegraphics[width=0.95\textwidth]{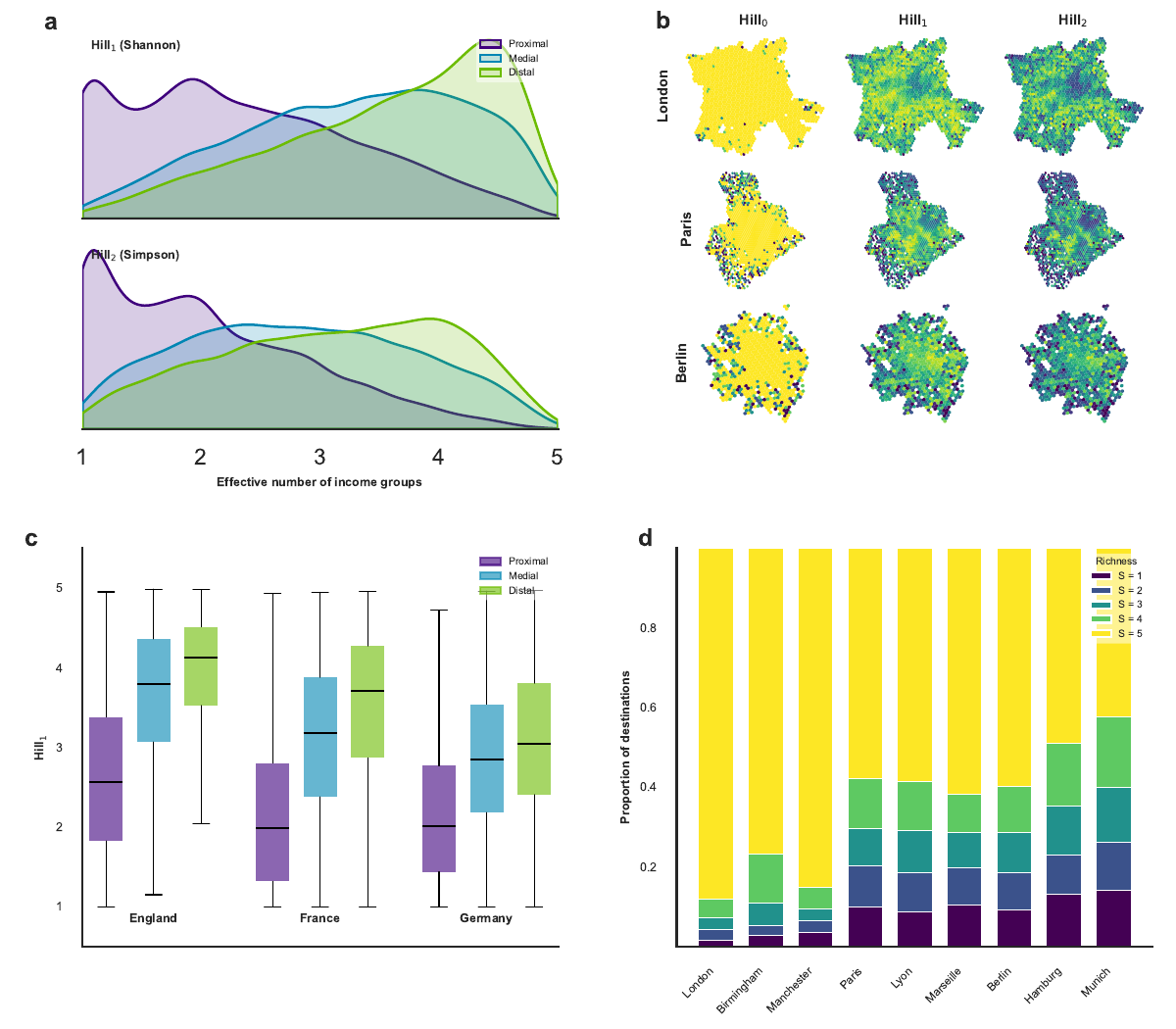}
\caption{\textbf{Distal trips broaden exposure without evening it.} \textbf{a} The distributions of Hill$_1$ and Hill$_2$ both climb from the proximal to the distal scale. \textbf{b} Maps of Hill$_0$, Hill$_1$, and Hill$_2$ across London, Paris, and Berlin trace the same centre--periphery gradient. \textbf{c} Hill$_1$ broken out by country and scale repeats the ordering everywhere. \textbf{d} The composition of destinations reached at each scale, city by city. Distal mobility broadens the range of groups encountered while leaving the evenness of exposure, $E_{12}=\mathrm{Hill}_2/\mathrm{Hill}_1$, largely unchanged.}
\label{si_diversity}
\end{figure*}

\clearpage

\section{Medialisation: variety, not reach}\label{si_medialisation_sec}

The medial scale recovers most of the mixing the longest trips provide, which raises a planning question: could a city route its long trips to nearby secondary centres and keep the mixing while shortening the journey? We test this with a deterministic counterfactual that medialises the distal component, reassigning each neighbourhood's distal trips to the city's subcentres rather than to its single core.

We read the subcentres from the visit surface itself. Counting visits to each \texttt{H3} level-7 hexagon gives a topography whose global peak is the nucleus; its remaining local peaks, taken greedily so that no two lie within five kilometres, are the subcentres. We reassign each neighbourhood's distal trips across them by a gravity rule that sends more trips to larger and nearer centres,
\[
P(k \mid i) \;\propto\; m_k \, \exp(-\beta\, d_{ik}),
\tag{S11}
\]
where $m_k$ is subcentre $k$'s visit mass and $d_{ik}$ its distance from neighbourhood $i$. We calibrate $\beta$ in each city so that the reassigned trips are, on average, as long as its observed medial trips, and hold the proximal and medial trips fixed. The regime is deterministic: we evaluate the resulting destination mix analytically, without sampling.

\begin{figure*}[!ht]
\centering
\includegraphics[width=0.95\textwidth]{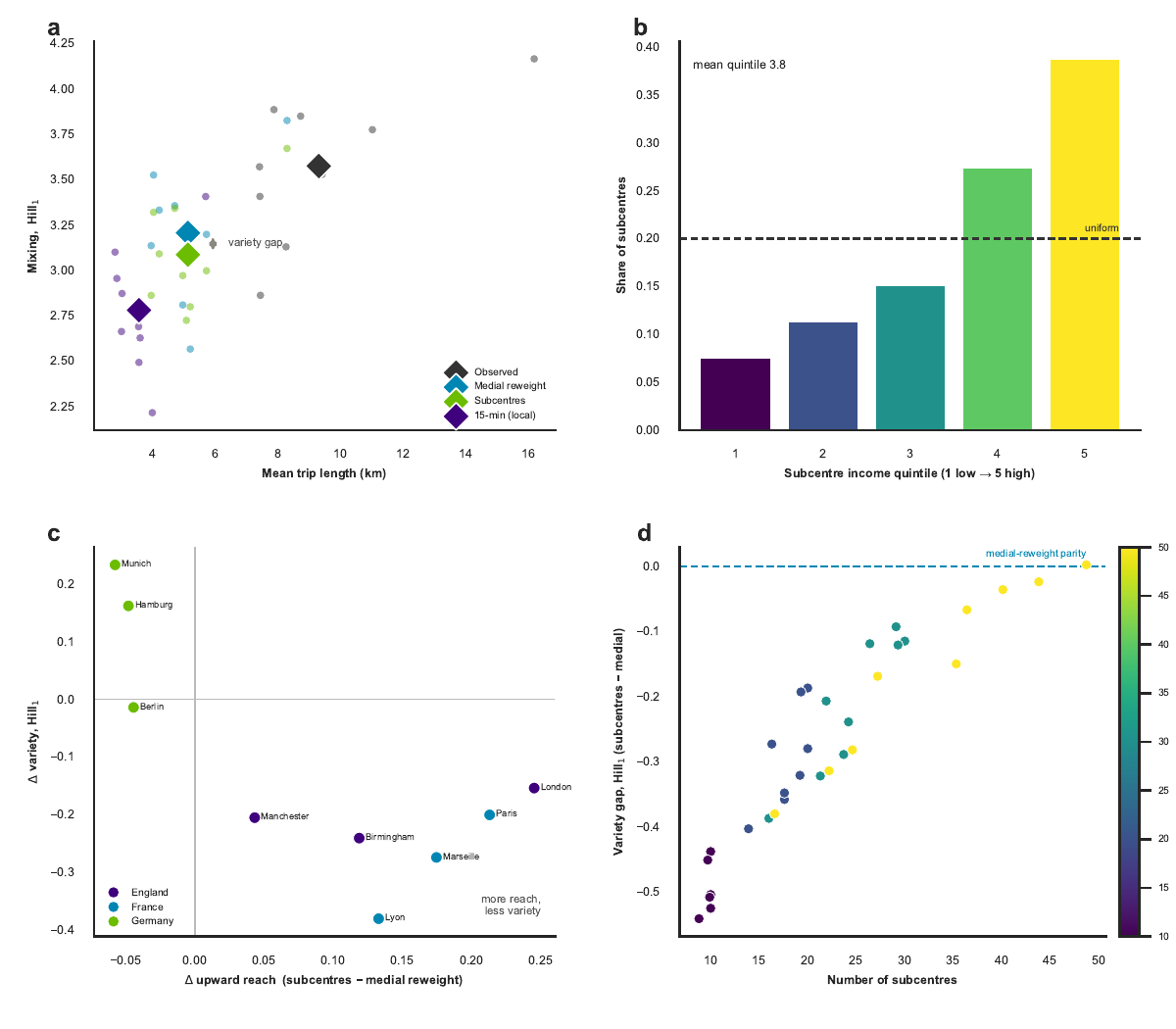}
\caption{\textbf{Medialisation trades variety for reach.} We reassign each neighbourhood's distal trips to the city's subcentres---the secondary peaks of the visit surface---by a size and distance gravity, holding shorter trips fixed and calibrating the reassigned length to the observed medial scale. \textbf{a} The move is efficient: it shortens the average trip by about 45\% and keeps mixing above the 15-minute floor in every city, but at the same length it mixes a little less than reweighting toward the diffuse medial---the variety gap. \textbf{b} Subcentres sit in the wealthier parts of the city, over-representing the top quintiles by area standing. \textbf{c} Their location is why: routing long trips to subcentres keeps, or raises, how far up the income ladder residents reach, but lowers the variety of areas they reach---more reach, less variety, in seven of nine cities. \textbf{d} The gap is robust across the subcentre definition, approaching the diffuse-medial spread only as hubs multiply. The medial scale's mixing comes from the variety of intermediate destinations, not from reaching higher.}
\label{si_medialisation}
\end{figure*}

The trade is real but one-sided. As Supplementary Fig.~\ref{si_medialisation}\textbf{a} shows, medialising to subcentres shortens the average trip by about 45\% and keeps mixing above the proximal floor in every city, yet at the same length it mixes a little less than simply reweighting the distal mass onto the medial scale. The reason lies in the centres themselves: Supplementary Fig.~\ref{si_medialisation}\textbf{b} shows that subcentres sit in the wealthier parts of the city, over-representing the top quintiles and barely touching the bottom. Routing residents there keeps, or even raises, how far up the income ladder they reach while narrowing the social range of the places they reach---the decomposition in Supplementary Fig.~\ref{si_medialisation}\textbf{c} makes the split plain: more reach, less variety, in seven of the nine cities. We score a destination by where it sits, not by who visits it, and these high-traffic hubs in fact draw a mixed crowd---each is visited by residents of every income quintile---so the narrowing falls on the range of \emph{areas} a resident's trips span, not on the company they keep on arrival. The gap is not an artefact of how we draw the subcentres; across thirty-six definitions that vary their number, spacing, and prominence, routing to subcentres never exceeds the diffuse medial spread, approaching it only as the centres multiply, which we show in Supplementary Fig.~\ref{si_medialisation}\textbf{d}.

The medial scale therefore mixes through the variety of places it reaches, not by reaching higher. A polycentric city built on its existing secondary centres would buy efficiency and preserve the upward pull toward wealthier districts, but would not broaden the range of places its residents reach; for that, the secondary centres would themselves have to be socioeconomically varied.

\end{document}